\documentclass[twocolumn]{aastex62}

\usepackage{bm}
\usepackage{amsmath,amsfonts,amssymb}
\usepackage{color}

\usepackage[draft]{minted}
\usepackage{graphicx}

\makeatletter
\def\PYG@reset{\let\PYG@it=\relax \let\PYG@bf=\relax%
    \let\PYG@ul=\relax \let\PYG@tc=\relax%
    \let\PYG@bc=\relax \let\PYG@ff=\relax}
\def\PYG@tok#1{\csname PYG@tok@#1\endcsname}
\def\PYG@toks#1+{\ifx\relax#1\empty\else%
    \PYG@tok{#1}\expandafter\PYG@toks\fi}
\def\PYG@do#1{\PYG@bc{\PYG@tc{\PYG@ul{%
    \PYG@it{\PYG@bf{\PYG@ff{#1}}}}}}}
\def\PYG#1#2{\PYG@reset\PYG@toks#1+\relax+\PYG@do{#2}}

\expandafter\def\csname PYG@tok@w\endcsname{\def\PYG@tc##1{\textcolor[rgb]{0.73,0.73,0.73}{##1}}}
\expandafter\def\csname PYG@tok@c\endcsname{\let\PYG@it=\textit\def\PYG@tc##1{\textcolor[rgb]{0.25,0.50,0.50}{##1}}}
\expandafter\def\csname PYG@tok@cp\endcsname{\def\PYG@tc##1{\textcolor[rgb]{0.74,0.48,0.00}{##1}}}
\expandafter\def\csname PYG@tok@k\endcsname{\let\PYG@bf=\textbf\def\PYG@tc##1{\textcolor[rgb]{0.00,0.50,0.00}{##1}}}
\expandafter\def\csname PYG@tok@kp\endcsname{\def\PYG@tc##1{\textcolor[rgb]{0.00,0.50,0.00}{##1}}}
\expandafter\def\csname PYG@tok@kt\endcsname{\def\PYG@tc##1{\textcolor[rgb]{0.69,0.00,0.25}{##1}}}
\expandafter\def\csname PYG@tok@o\endcsname{\def\PYG@tc##1{\textcolor[rgb]{0.40,0.40,0.40}{##1}}}
\expandafter\def\csname PYG@tok@ow\endcsname{\let\PYG@bf=\textbf\def\PYG@tc##1{\textcolor[rgb]{0.67,0.13,1.00}{##1}}}
\expandafter\def\csname PYG@tok@nb\endcsname{\def\PYG@tc##1{\textcolor[rgb]{0.00,0.50,0.00}{##1}}}
\expandafter\def\csname PYG@tok@nf\endcsname{\def\PYG@tc##1{\textcolor[rgb]{0.00,0.00,1.00}{##1}}}
\expandafter\def\csname PYG@tok@nc\endcsname{\let\PYG@bf=\textbf\def\PYG@tc##1{\textcolor[rgb]{0.00,0.00,1.00}{##1}}}
\expandafter\def\csname PYG@tok@nn\endcsname{\let\PYG@bf=\textbf\def\PYG@tc##1{\textcolor[rgb]{0.00,0.00,1.00}{##1}}}
\expandafter\def\csname PYG@tok@ne\endcsname{\let\PYG@bf=\textbf\def\PYG@tc##1{\textcolor[rgb]{0.82,0.25,0.23}{##1}}}
\expandafter\def\csname PYG@tok@nv\endcsname{\def\PYG@tc##1{\textcolor[rgb]{0.10,0.09,0.49}{##1}}}
\expandafter\def\csname PYG@tok@no\endcsname{\def\PYG@tc##1{\textcolor[rgb]{0.53,0.00,0.00}{##1}}}
\expandafter\def\csname PYG@tok@nl\endcsname{\def\PYG@tc##1{\textcolor[rgb]{0.63,0.63,0.00}{##1}}}
\expandafter\def\csname PYG@tok@ni\endcsname{\let\PYG@bf=\textbf\def\PYG@tc##1{\textcolor[rgb]{0.60,0.60,0.60}{##1}}}
\expandafter\def\csname PYG@tok@na\endcsname{\def\PYG@tc##1{\textcolor[rgb]{0.49,0.56,0.16}{##1}}}
\expandafter\def\csname PYG@tok@nt\endcsname{\let\PYG@bf=\textbf\def\PYG@tc##1{\textcolor[rgb]{0.00,0.50,0.00}{##1}}}
\expandafter\def\csname PYG@tok@nd\endcsname{\def\PYG@tc##1{\textcolor[rgb]{0.67,0.13,1.00}{##1}}}
\expandafter\def\csname PYG@tok@s\endcsname{\def\PYG@tc##1{\textcolor[rgb]{0.73,0.13,0.13}{##1}}}
\expandafter\def\csname PYG@tok@sd\endcsname{\let\PYG@it=\textit\def\PYG@tc##1{\textcolor[rgb]{0.73,0.13,0.13}{##1}}}
\expandafter\def\csname PYG@tok@si\endcsname{\let\PYG@bf=\textbf\def\PYG@tc##1{\textcolor[rgb]{0.73,0.40,0.53}{##1}}}
\expandafter\def\csname PYG@tok@se\endcsname{\let\PYG@bf=\textbf\def\PYG@tc##1{\textcolor[rgb]{0.73,0.40,0.13}{##1}}}
\expandafter\def\csname PYG@tok@sr\endcsname{\def\PYG@tc##1{\textcolor[rgb]{0.73,0.40,0.53}{##1}}}
\expandafter\def\csname PYG@tok@ss\endcsname{\def\PYG@tc##1{\textcolor[rgb]{0.10,0.09,0.49}{##1}}}
\expandafter\def\csname PYG@tok@sx\endcsname{\def\PYG@tc##1{\textcolor[rgb]{0.00,0.50,0.00}{##1}}}
\expandafter\def\csname PYG@tok@m\endcsname{\def\PYG@tc##1{\textcolor[rgb]{0.40,0.40,0.40}{##1}}}
\expandafter\def\csname PYG@tok@gh\endcsname{\let\PYG@bf=\textbf\def\PYG@tc##1{\textcolor[rgb]{0.00,0.00,0.50}{##1}}}
\expandafter\def\csname PYG@tok@gu\endcsname{\let\PYG@bf=\textbf\def\PYG@tc##1{\textcolor[rgb]{0.50,0.00,0.50}{##1}}}
\expandafter\def\csname PYG@tok@gd\endcsname{\def\PYG@tc##1{\textcolor[rgb]{0.63,0.00,0.00}{##1}}}
\expandafter\def\csname PYG@tok@gi\endcsname{\def\PYG@tc##1{\textcolor[rgb]{0.00,0.63,0.00}{##1}}}
\expandafter\def\csname PYG@tok@gr\endcsname{\def\PYG@tc##1{\textcolor[rgb]{1.00,0.00,0.00}{##1}}}
\expandafter\def\csname PYG@tok@ge\endcsname{\let\PYG@it=\textit}
\expandafter\def\csname PYG@tok@gs\endcsname{\let\PYG@bf=\textbf}
\expandafter\def\csname PYG@tok@gp\endcsname{\let\PYG@bf=\textbf\def\PYG@tc##1{\textcolor[rgb]{0.00,0.00,0.50}{##1}}}
\expandafter\def\csname PYG@tok@go\endcsname{\def\PYG@tc##1{\textcolor[rgb]{0.53,0.53,0.53}{##1}}}
\expandafter\def\csname PYG@tok@gt\endcsname{\def\PYG@tc##1{\textcolor[rgb]{0.00,0.27,0.87}{##1}}}
\expandafter\def\csname PYG@tok@err\endcsname{\def\PYG@bc##1{\setlength{\fboxsep}{0pt}\fcolorbox[rgb]{1.00,0.00,0.00}{1,1,1}{\strut ##1}}}
\expandafter\def\csname PYG@tok@kc\endcsname{\let\PYG@bf=\textbf\def\PYG@tc##1{\textcolor[rgb]{0.00,0.50,0.00}{##1}}}
\expandafter\def\csname PYG@tok@kd\endcsname{\let\PYG@bf=\textbf\def\PYG@tc##1{\textcolor[rgb]{0.00,0.50,0.00}{##1}}}
\expandafter\def\csname PYG@tok@kn\endcsname{\let\PYG@bf=\textbf\def\PYG@tc##1{\textcolor[rgb]{0.00,0.50,0.00}{##1}}}
\expandafter\def\csname PYG@tok@kr\endcsname{\let\PYG@bf=\textbf\def\PYG@tc##1{\textcolor[rgb]{0.00,0.50,0.00}{##1}}}
\expandafter\def\csname PYG@tok@bp\endcsname{\def\PYG@tc##1{\textcolor[rgb]{0.00,0.50,0.00}{##1}}}
\expandafter\def\csname PYG@tok@fm\endcsname{\def\PYG@tc##1{\textcolor[rgb]{0.00,0.00,1.00}{##1}}}
\expandafter\def\csname PYG@tok@vc\endcsname{\def\PYG@tc##1{\textcolor[rgb]{0.10,0.09,0.49}{##1}}}
\expandafter\def\csname PYG@tok@vg\endcsname{\def\PYG@tc##1{\textcolor[rgb]{0.10,0.09,0.49}{##1}}}
\expandafter\def\csname PYG@tok@vi\endcsname{\def\PYG@tc##1{\textcolor[rgb]{0.10,0.09,0.49}{##1}}}
\expandafter\def\csname PYG@tok@vm\endcsname{\def\PYG@tc##1{\textcolor[rgb]{0.10,0.09,0.49}{##1}}}
\expandafter\def\csname PYG@tok@sa\endcsname{\def\PYG@tc##1{\textcolor[rgb]{0.73,0.13,0.13}{##1}}}
\expandafter\def\csname PYG@tok@sb\endcsname{\def\PYG@tc##1{\textcolor[rgb]{0.73,0.13,0.13}{##1}}}
\expandafter\def\csname PYG@tok@sc\endcsname{\def\PYG@tc##1{\textcolor[rgb]{0.73,0.13,0.13}{##1}}}
\expandafter\def\csname PYG@tok@dl\endcsname{\def\PYG@tc##1{\textcolor[rgb]{0.73,0.13,0.13}{##1}}}
\expandafter\def\csname PYG@tok@s2\endcsname{\def\PYG@tc##1{\textcolor[rgb]{0.73,0.13,0.13}{##1}}}
\expandafter\def\csname PYG@tok@sh\endcsname{\def\PYG@tc##1{\textcolor[rgb]{0.73,0.13,0.13}{##1}}}
\expandafter\def\csname PYG@tok@s1\endcsname{\def\PYG@tc##1{\textcolor[rgb]{0.73,0.13,0.13}{##1}}}
\expandafter\def\csname PYG@tok@mb\endcsname{\def\PYG@tc##1{\textcolor[rgb]{0.40,0.40,0.40}{##1}}}
\expandafter\def\csname PYG@tok@mf\endcsname{\def\PYG@tc##1{\textcolor[rgb]{0.40,0.40,0.40}{##1}}}
\expandafter\def\csname PYG@tok@mh\endcsname{\def\PYG@tc##1{\textcolor[rgb]{0.40,0.40,0.40}{##1}}}
\expandafter\def\csname PYG@tok@mi\endcsname{\def\PYG@tc##1{\textcolor[rgb]{0.40,0.40,0.40}{##1}}}
\expandafter\def\csname PYG@tok@il\endcsname{\def\PYG@tc##1{\textcolor[rgb]{0.40,0.40,0.40}{##1}}}
\expandafter\def\csname PYG@tok@mo\endcsname{\def\PYG@tc##1{\textcolor[rgb]{0.40,0.40,0.40}{##1}}}
\expandafter\def\csname PYG@tok@ch\endcsname{\let\PYG@it=\textit\def\PYG@tc##1{\textcolor[rgb]{0.25,0.50,0.50}{##1}}}
\expandafter\def\csname PYG@tok@cm\endcsname{\let\PYG@it=\textit\def\PYG@tc##1{\textcolor[rgb]{0.25,0.50,0.50}{##1}}}
\expandafter\def\csname PYG@tok@cpf\endcsname{\let\PYG@it=\textit\def\PYG@tc##1{\textcolor[rgb]{0.25,0.50,0.50}{##1}}}
\expandafter\def\csname PYG@tok@c1\endcsname{\let\PYG@it=\textit\def\PYG@tc##1{\textcolor[rgb]{0.25,0.50,0.50}{##1}}}
\expandafter\def\csname PYG@tok@cs\endcsname{\let\PYG@it=\textit\def\PYG@tc##1{\textcolor[rgb]{0.25,0.50,0.50}{##1}}}

\makeatother

\makeatletter
\def\PYGdefault@reset{\let\PYGdefault@it=\relax \let\PYGdefault@bf=\relax%
    \let\PYGdefault@ul=\relax \let\PYGdefault@tc=\relax%
    \let\PYGdefault@bc=\relax \let\PYGdefault@ff=\relax}
\def\PYGdefault@tok#1{\csname PYGdefault@tok@#1\endcsname}
\def\PYGdefault@toks#1+{\ifx\relax#1\empty\else%
    \PYGdefault@tok{#1}\expandafter\PYGdefault@toks\fi}
\def\PYGdefault@do#1{\PYGdefault@bc{\PYGdefault@tc{\PYGdefault@ul{%
    \PYGdefault@it{\PYGdefault@bf{\PYGdefault@ff{#1}}}}}}}
\def\PYGdefault#1#2{\PYGdefault@reset\PYGdefault@toks#1+\relax+\PYGdefault@do{#2}}

\expandafter\def\csname PYGdefault@tok@w\endcsname{\def\PYGdefault@tc##1{\textcolor[rgb]{0.73,0.73,0.73}{##1}}}
\expandafter\def\csname PYGdefault@tok@c\endcsname{\let\PYGdefault@it=\textit\def\PYGdefault@tc##1{\textcolor[rgb]{0.25,0.50,0.50}{##1}}}
\expandafter\def\csname PYGdefault@tok@cp\endcsname{\def\PYGdefault@tc##1{\textcolor[rgb]{0.74,0.48,0.00}{##1}}}
\expandafter\def\csname PYGdefault@tok@k\endcsname{\let\PYGdefault@bf=\textbf\def\PYGdefault@tc##1{\textcolor[rgb]{0.00,0.50,0.00}{##1}}}
\expandafter\def\csname PYGdefault@tok@kp\endcsname{\def\PYGdefault@tc##1{\textcolor[rgb]{0.00,0.50,0.00}{##1}}}
\expandafter\def\csname PYGdefault@tok@kt\endcsname{\def\PYGdefault@tc##1{\textcolor[rgb]{0.69,0.00,0.25}{##1}}}
\expandafter\def\csname PYGdefault@tok@o\endcsname{\def\PYGdefault@tc##1{\textcolor[rgb]{0.40,0.40,0.40}{##1}}}
\expandafter\def\csname PYGdefault@tok@ow\endcsname{\let\PYGdefault@bf=\textbf\def\PYGdefault@tc##1{\textcolor[rgb]{0.67,0.13,1.00}{##1}}}
\expandafter\def\csname PYGdefault@tok@nb\endcsname{\def\PYGdefault@tc##1{\textcolor[rgb]{0.00,0.50,0.00}{##1}}}
\expandafter\def\csname PYGdefault@tok@nf\endcsname{\def\PYGdefault@tc##1{\textcolor[rgb]{0.00,0.00,1.00}{##1}}}
\expandafter\def\csname PYGdefault@tok@nc\endcsname{\let\PYGdefault@bf=\textbf\def\PYGdefault@tc##1{\textcolor[rgb]{0.00,0.00,1.00}{##1}}}
\expandafter\def\csname PYGdefault@tok@nn\endcsname{\let\PYGdefault@bf=\textbf\def\PYGdefault@tc##1{\textcolor[rgb]{0.00,0.00,1.00}{##1}}}
\expandafter\def\csname PYGdefault@tok@ne\endcsname{\let\PYGdefault@bf=\textbf\def\PYGdefault@tc##1{\textcolor[rgb]{0.82,0.25,0.23}{##1}}}
\expandafter\def\csname PYGdefault@tok@nv\endcsname{\def\PYGdefault@tc##1{\textcolor[rgb]{0.10,0.09,0.49}{##1}}}
\expandafter\def\csname PYGdefault@tok@no\endcsname{\def\PYGdefault@tc##1{\textcolor[rgb]{0.53,0.00,0.00}{##1}}}
\expandafter\def\csname PYGdefault@tok@nl\endcsname{\def\PYGdefault@tc##1{\textcolor[rgb]{0.63,0.63,0.00}{##1}}}
\expandafter\def\csname PYGdefault@tok@ni\endcsname{\let\PYGdefault@bf=\textbf\def\PYGdefault@tc##1{\textcolor[rgb]{0.60,0.60,0.60}{##1}}}
\expandafter\def\csname PYGdefault@tok@na\endcsname{\def\PYGdefault@tc##1{\textcolor[rgb]{0.49,0.56,0.16}{##1}}}
\expandafter\def\csname PYGdefault@tok@nt\endcsname{\let\PYGdefault@bf=\textbf\def\PYGdefault@tc##1{\textcolor[rgb]{0.00,0.50,0.00}{##1}}}
\expandafter\def\csname PYGdefault@tok@nd\endcsname{\def\PYGdefault@tc##1{\textcolor[rgb]{0.67,0.13,1.00}{##1}}}
\expandafter\def\csname PYGdefault@tok@s\endcsname{\def\PYGdefault@tc##1{\textcolor[rgb]{0.73,0.13,0.13}{##1}}}
\expandafter\def\csname PYGdefault@tok@sd\endcsname{\let\PYGdefault@it=\textit\def\PYGdefault@tc##1{\textcolor[rgb]{0.73,0.13,0.13}{##1}}}
\expandafter\def\csname PYGdefault@tok@si\endcsname{\let\PYGdefault@bf=\textbf\def\PYGdefault@tc##1{\textcolor[rgb]{0.73,0.40,0.53}{##1}}}
\expandafter\def\csname PYGdefault@tok@se\endcsname{\let\PYGdefault@bf=\textbf\def\PYGdefault@tc##1{\textcolor[rgb]{0.73,0.40,0.13}{##1}}}
\expandafter\def\csname PYGdefault@tok@sr\endcsname{\def\PYGdefault@tc##1{\textcolor[rgb]{0.73,0.40,0.53}{##1}}}
\expandafter\def\csname PYGdefault@tok@ss\endcsname{\def\PYGdefault@tc##1{\textcolor[rgb]{0.10,0.09,0.49}{##1}}}
\expandafter\def\csname PYGdefault@tok@sx\endcsname{\def\PYGdefault@tc##1{\textcolor[rgb]{0.00,0.50,0.00}{##1}}}
\expandafter\def\csname PYGdefault@tok@m\endcsname{\def\PYGdefault@tc##1{\textcolor[rgb]{0.40,0.40,0.40}{##1}}}
\expandafter\def\csname PYGdefault@tok@gh\endcsname{\let\PYGdefault@bf=\textbf\def\PYGdefault@tc##1{\textcolor[rgb]{0.00,0.00,0.50}{##1}}}
\expandafter\def\csname PYGdefault@tok@gu\endcsname{\let\PYGdefault@bf=\textbf\def\PYGdefault@tc##1{\textcolor[rgb]{0.50,0.00,0.50}{##1}}}
\expandafter\def\csname PYGdefault@tok@gd\endcsname{\def\PYGdefault@tc##1{\textcolor[rgb]{0.63,0.00,0.00}{##1}}}
\expandafter\def\csname PYGdefault@tok@gi\endcsname{\def\PYGdefault@tc##1{\textcolor[rgb]{0.00,0.63,0.00}{##1}}}
\expandafter\def\csname PYGdefault@tok@gr\endcsname{\def\PYGdefault@tc##1{\textcolor[rgb]{1.00,0.00,0.00}{##1}}}
\expandafter\def\csname PYGdefault@tok@ge\endcsname{\let\PYGdefault@it=\textit}
\expandafter\def\csname PYGdefault@tok@gs\endcsname{\let\PYGdefault@bf=\textbf}
\expandafter\def\csname PYGdefault@tok@gp\endcsname{\let\PYGdefault@bf=\textbf\def\PYGdefault@tc##1{\textcolor[rgb]{0.00,0.00,0.50}{##1}}}
\expandafter\def\csname PYGdefault@tok@go\endcsname{\def\PYGdefault@tc##1{\textcolor[rgb]{0.53,0.53,0.53}{##1}}}
\expandafter\def\csname PYGdefault@tok@gt\endcsname{\def\PYGdefault@tc##1{\textcolor[rgb]{0.00,0.27,0.87}{##1}}}
\expandafter\def\csname PYGdefault@tok@err\endcsname{\def\PYGdefault@bc##1{\setlength{\fboxsep}{0pt}\fcolorbox[rgb]{1.00,0.00,0.00}{1,1,1}{\strut ##1}}}
\expandafter\def\csname PYGdefault@tok@kc\endcsname{\let\PYGdefault@bf=\textbf\def\PYGdefault@tc##1{\textcolor[rgb]{0.00,0.50,0.00}{##1}}}
\expandafter\def\csname PYGdefault@tok@kd\endcsname{\let\PYGdefault@bf=\textbf\def\PYGdefault@tc##1{\textcolor[rgb]{0.00,0.50,0.00}{##1}}}
\expandafter\def\csname PYGdefault@tok@kn\endcsname{\let\PYGdefault@bf=\textbf\def\PYGdefault@tc##1{\textcolor[rgb]{0.00,0.50,0.00}{##1}}}
\expandafter\def\csname PYGdefault@tok@kr\endcsname{\let\PYGdefault@bf=\textbf\def\PYGdefault@tc##1{\textcolor[rgb]{0.00,0.50,0.00}{##1}}}
\expandafter\def\csname PYGdefault@tok@bp\endcsname{\def\PYGdefault@tc##1{\textcolor[rgb]{0.00,0.50,0.00}{##1}}}
\expandafter\def\csname PYGdefault@tok@fm\endcsname{\def\PYGdefault@tc##1{\textcolor[rgb]{0.00,0.00,1.00}{##1}}}
\expandafter\def\csname PYGdefault@tok@vc\endcsname{\def\PYGdefault@tc##1{\textcolor[rgb]{0.10,0.09,0.49}{##1}}}
\expandafter\def\csname PYGdefault@tok@vg\endcsname{\def\PYGdefault@tc##1{\textcolor[rgb]{0.10,0.09,0.49}{##1}}}
\expandafter\def\csname PYGdefault@tok@vi\endcsname{\def\PYGdefault@tc##1{\textcolor[rgb]{0.10,0.09,0.49}{##1}}}
\expandafter\def\csname PYGdefault@tok@vm\endcsname{\def\PYGdefault@tc##1{\textcolor[rgb]{0.10,0.09,0.49}{##1}}}
\expandafter\def\csname PYGdefault@tok@sa\endcsname{\def\PYGdefault@tc##1{\textcolor[rgb]{0.73,0.13,0.13}{##1}}}
\expandafter\def\csname PYGdefault@tok@sb\endcsname{\def\PYGdefault@tc##1{\textcolor[rgb]{0.73,0.13,0.13}{##1}}}
\expandafter\def\csname PYGdefault@tok@sc\endcsname{\def\PYGdefault@tc##1{\textcolor[rgb]{0.73,0.13,0.13}{##1}}}
\expandafter\def\csname PYGdefault@tok@dl\endcsname{\def\PYGdefault@tc##1{\textcolor[rgb]{0.73,0.13,0.13}{##1}}}
\expandafter\def\csname PYGdefault@tok@s2\endcsname{\def\PYGdefault@tc##1{\textcolor[rgb]{0.73,0.13,0.13}{##1}}}
\expandafter\def\csname PYGdefault@tok@sh\endcsname{\def\PYGdefault@tc##1{\textcolor[rgb]{0.73,0.13,0.13}{##1}}}
\expandafter\def\csname PYGdefault@tok@s1\endcsname{\def\PYGdefault@tc##1{\textcolor[rgb]{0.73,0.13,0.13}{##1}}}
\expandafter\def\csname PYGdefault@tok@mb\endcsname{\def\PYGdefault@tc##1{\textcolor[rgb]{0.40,0.40,0.40}{##1}}}
\expandafter\def\csname PYGdefault@tok@mf\endcsname{\def\PYGdefault@tc##1{\textcolor[rgb]{0.40,0.40,0.40}{##1}}}
\expandafter\def\csname PYGdefault@tok@mh\endcsname{\def\PYGdefault@tc##1{\textcolor[rgb]{0.40,0.40,0.40}{##1}}}
\expandafter\def\csname PYGdefault@tok@mi\endcsname{\def\PYGdefault@tc##1{\textcolor[rgb]{0.40,0.40,0.40}{##1}}}
\expandafter\def\csname PYGdefault@tok@il\endcsname{\def\PYGdefault@tc##1{\textcolor[rgb]{0.40,0.40,0.40}{##1}}}
\expandafter\def\csname PYGdefault@tok@mo\endcsname{\def\PYGdefault@tc##1{\textcolor[rgb]{0.40,0.40,0.40}{##1}}}
\expandafter\def\csname PYGdefault@tok@ch\endcsname{\let\PYGdefault@it=\textit\def\PYGdefault@tc##1{\textcolor[rgb]{0.25,0.50,0.50}{##1}}}
\expandafter\def\csname PYGdefault@tok@cm\endcsname{\let\PYGdefault@it=\textit\def\PYGdefault@tc##1{\textcolor[rgb]{0.25,0.50,0.50}{##1}}}
\expandafter\def\csname PYGdefault@tok@cpf\endcsname{\let\PYGdefault@it=\textit\def\PYGdefault@tc##1{\textcolor[rgb]{0.25,0.50,0.50}{##1}}}
\expandafter\def\csname PYGdefault@tok@c1\endcsname{\let\PYGdefault@it=\textit\def\PYGdefault@tc##1{\textcolor[rgb]{0.25,0.50,0.50}{##1}}}
\expandafter\def\csname PYGdefault@tok@cs\endcsname{\let\PYGdefault@it=\textit\def\PYGdefault@tc##1{\textcolor[rgb]{0.25,0.50,0.50}{##1}}}

\makeatother

\newcommand\teff{T_{\rm eff}}
\newcommand\logg{\log{g}}

\newcommand{\project}[1]{\textsl{#1}}
\newcommand{\package}[1]{\texttt{#1}}

\newcommand{\Galah}{\project{Galah}}

\newcommand{\vect}[1]{\boldsymbol{\mathbf{#1}}}
\renewcommand{\vec}[1]{\vect{#1}}

\newcommand{\weight}{\pi}
\newcommand{\data}{\textbf{Y}}
\newcommand{\vecdata}{\vec\data}
\newcommand{\vecdataunscaled}{\vec{X}}
\newcommand{\diag}[1]{\textrm{diag}(#1)}

\newcommand{\nextstep}{^\textrm{(t+1)}}
\newcommand{\thisstep}{^\textrm{(t)}}
\newcommand{\transpose}{^\intercal}
\newcommand{\eye}{\textbf{I}}

\newcommand{\factorloads}{\textbf{L}}
\newcommand{\factorscores}{\textbf{S}}
\newcommand{\specificvariance}{\vec{D}}

\newcommand{\scoremeans}{\vec\xi}
\newcommand{\scorecovs}{\vec\Omega}

\newcommand{\NumData}{N}
\newcommand{\NumDimensions}{D}
\newcommand{\numdata}{n}

\newcommand{\NumLatentFactors}{J}
\newcommand{\numlatentfactors}{j}
\newcommand{\NumComponents}{K}
\newcommand{\numcomponents}{k}
\newcommand{\likelihood}{\mathcal{L}}

\newcommand{\EvaluationHash}{c7d68}

\received{2019}
\revised{2019}
\accepted{2019}

\newcommand{\vcpath}{vc.tex}

\IfFileExists{\vcpath}{%%% This file is generated by the Makefile.

}{

}

\submitjournal{AAS Journals}

\shorttitle{A data-driven model of nucleosynthesis with chemical tagging in latent space}
\shortauthors{Casey et al.}

\begin{document}

\title{A data-driven model of nucleosynthesis with chemical tagging in a lower-dimensional latent space}

\correspondingauthor{Andrew R. Casey}
\email{andrew.casey@monash.edu}

\author[0000-0003-0174-0564]{Andrew R. Casey}
\affiliation{School of Physics \& Astronomy, 
			 Monash University,
			 Wellington Rd, Clayton 3800, Victoria, Australia}
\affiliation{Faculty of Information Technology, 
			 Monash University, 
			 Wellington Rd, Clayton 3800, Victoria, Australia}
			 
\author[0000-0003-2952-859X]{John C. Lattanzio}
\affiliation{School of Physics \& Astronomy, 
			 Monash University,
			 Wellington Rd, Clayton 3800, Victoria, Australia}

\author[0000-0002-1716-690X]{Aldeida Aleti}
\affiliation{Faculty of Information Technology, 
			 Monash University, 
			 Wellington Rd, Clayton 3800, Victoria, Australia}

\author[0000-0002-0583-5918]{David L. Dowe}
\affiliation{Faculty of Information Technology, 
			 Monash University, 
			 Wellington Rd, Clayton 3800, Victoria, Australia}

\author[0000-0001-7516-4016]{Joss Bland-Hawthorn}
\affiliation{Sydney Institute for Astronomy, School of Physics,
			 A28, The University of Sydney, NSW 2006, Australia}
\affiliation{Center of Excellence for Astrophysics in Three Dimensions (ASTRO-3D),
			 Australia}
\affiliation{Miller Professor, Miller Institute, 
			 UC Berkeley, 
			 Berkeley, CA 94720, USA}

\author[0000-0002-4031-8553]{Sven Buder}
\affiliation{Research School of Astronomy and Astrophysics,
			 Australian National University,
			 Canberra, ACT 2611, Australia}

\author[0000-0003-3081-9319]{Geraint F. Lewis}
\affiliation{Sydney Institute for Astronomy, School of Physics,
			 A28, The University of Sydney, NSW 2006, Australia}

\author[0000-0002-3430-4163]{Sarah L. Martell}
\affiliation{School of Physics,
			 University of New South Wales, 
			 Sydney, NSW 2052, Australia}
	
\author[0000-0001-5344-8069]{Thomas Nordlander}			 
\affiliation{Research School of Astronomy and Astrophysics,
			 Australian National University,
			 Canberra, ACT 2611, Australia}
\affiliation{Center of Excellence for Astrophysics in Three Dimensions (ASTRO-3D), 
			 Australia}
			 
\author[0000-0002-8165-2507]{Jeffrey D. Simpson}
\affiliation{School of Physics,
			 University of New South Wales, 
			 Sydney, NSW 2052, Australia}

\author[0000-0002-0920-809X]{\added{Sanjib Sharma}}
\affiliation{Sydney Institute for Astronomy, School of Physics,
			 A28, The University of Sydney, NSW 2006, Australia}
						 
\author[0000-0003-1124-8477]{Daniel B. Zucker}
\affiliation{Department of Physics and Astronomy,
			 Macquarie University, 
			 Sydney, NSW 2109, Australia}

\begin{abstract}
Chemical tagging seeks to identify unique star formation sites from
present-day stellar abundances.
Previous techniques have treated each abundance dimension as being
statistically independent, despite theoretical expectations that
many elements can be produced by more than one nucleosynthetic process.
In this work we introduce a data-driven model of nucleosynthesis 
where a set of latent factors (e.g., nucleosynthetic yields) contribute
to all stars with different scores, and clustering (e.g., chemical tagging) 
is modelled by a mixture of multivariate Gaussians
in a lower-dimensional latent space.
We use an exact method to simultaneously estimate the factor scores for
each star, the partial assignment of each star
to each cluster, and the latent factors common to all stars, even in the
presence of missing data entries.
We use an information-theoretic Bayesian principle to estimate the number of
latent factors and clusters.
Using the second \Galah\ data release we find that six latent factors are
preferred to explain $N =$ \replaced{1,072}{2,566} stars with 17 chemical abundances.
We identify the rapid- and slow-neutron capture 
processes, as well as latent factors consistent with Fe-peak and 
$\alpha$-element production, and another where K and Zn dominate.
When we consider $N \sim$ \replaced{100,000}{160,000} stars with missing abundances we find 
another \replaced{five}{seven} factors, as well as 16 components in latent space.
Despite these components showing separation in chemistry that is explained
through different yield contributions, none show significant structure in 
their positions or motions. We argue that more data, and joint priors 
on cluster membership that are constrained by dynamical models, are necessary
to realise chemical tagging at a galactic-scale.
We release accompanying software that scales well with the available data,
allowing for model parameters to be optimised in seconds given a fixed number
of latent factors, components, and $\sim10^7$ abundance measurements.
\end{abstract}

\keywords{Bayesian statistics (1900), Chemical abundances (224), Galaxy chemical evolution (580)}

\section{Introduction} \label{sec:intro}

The detailed chemical abundances that are observable in a star's photosphere provide a
fossil record that carries with it information about where and when that star
formed \citep{Freeman;Bland-Hawthorn:2002}. While the photospheric abundances remain largely unchanged throughout
a star's lifetime \citep[however see][]{Dotter:2017,Ness:2018b}, the dynamical 
dissipation timescale of open clusters in the Milky Way disc is of order a few 
gigayears \citep{Portegies-Zwart:1998}. That makes chemical tagging an attractive 
approach to identify star formation sites long after those stars are no longer 
gravitationally bound to each other.

Gravitationally bound star clusters have been useful laboratories for
testing the limits and utility of chemical tagging. Although biases arise when
only considering star clusters that are still gravitationally bound, the chemical
homogeneity of open clusters provides an empirical measure of how similar stars
would need to be before they could be tagged as belonging to the same
star formation site \citep{Bland-Hawthorn:2010b,Bland-Hawthorn:2010a,Mitschang:2014}. However, there are analysis
issues in understanding how precisely those chemical abundances can be measured
\citep{Bovy:2016}, and how chemically similar stars can be that did not form 
together \citep[doppleg\"angers;][]{Ness:2018}.
If open clusters were truely chemically homogeneous then under idealistic 
assumptions our ability to chemically tag the Milky Way would depend primarily
on the precision with which we can measure those chemical abundances in stars. 
Data-driven approaches to modelling stellar spectra are
improving upon this precision \citep{Ness:2015,Ness:2018a,Ness:2018b,
Casey:2016,Casey:2017,Ho:2017b,Ho:2017a,Leung;Bovy:2018,Ting:2019}, but more work is
needed: astronomers have not yet developed unbiased estimators of chemical
abundances that saturate the Cram\'er-Rao bound \citep{Cramer:1946,Rao:1945}.

Chemical tagging experiments require a catalogue of precise chemical abundance
measurements for a large number of stars,
where those chemical abundances trace different nucleosynthetic pathways.
This is the primary goal of the Galactic Archaeology with \project{HERMES} (\Galah) survey \citep{DeSilva:2015,Martell:2017,Buder:2018},
a stellar spectroscopic survey that uses the High Efficiency and Resolution 
Multi-Element Spectrograph \citep[\project{HERMES};][]{Sheinis:2015} on the 3.9~m Anglo-Australian 
Telescope (AAT).  \Galah\ will observe up to $10^6$ stars in the 
Milky Way, and measure up to 30 chemical abundances for each star \citep{Bland-Hawthorn:2016}. This includes
light odd-Z elements (e.g., Na, K), elements produced through
alpha-particle capture (e.g., Mg, Ca\deleted{, Ti}), and elements produced
through the slow (e.g., Ba) and rapid neutron-capture process
(e.g., Eu). No other current or planned spectroscopic survey provides an equivalent set of
chemical abundances for a comparable number of stars.

Given these data and the most favourable assumptions in chemical tagging 
-- that star clusters are truely chemically homogenous, that we can measure 
those abundances with infinite precision, and that those abundances are 
differentiable between star clusters -- then chemical
tagging becomes a clustering problem. All clustering techniques applied to 
chemical tagging thus far have assumed that the data dimensions are independent. That is to say
that adding a dimension of say [Ni/H] provides independent information
that could not have been predicted from other elemental abundances.
Theory and observations agree that this cannot be true.
Nucleosynthetic processes produce multiple elements in varying
quantities, and the effective dimensionality of stellar abundance datasets has been shown
to be lower than the actual number of abundance dimensions \citep{Ting:2012,Price-Jones:2018,Milosavljevic:2018}.
Any clustering approach that treats each new elemental abundance as an 
independent axis of information will therefore conclude with biased inferences
about the star formation history of our Galaxy.

It is not trivial to confidently estimate
the nucleosynthetic yields that have contributed to the chemical abundances of each star. There are
qualitative statements that can be made for large numbers of stars, or particular
types of stars, but quantifying the precise contribution of different processes
to each star is an unsolved problem. For example, the so-called [$\alpha$/Fe] `knee' in
abundance ratios in the Milky Way can qualitatively be explained by 
core-collapse supernovae being the predominant nucleosynthetic process in the
early Milky Way before Type Ia supernovae made a significant contribution, but 
efforts to date have not sought to try to explain the detailed abundances of 
stars as a contribution of yields from different systems \citep[however see][]{West:2013}.
This is in part 
because of the challenging and degenerate nature of the problem as described, 
and is complicated by the differences in yield predictions that account from 
prescriptions used in different theoretical models.

New approaches to chemical tagging are clearly needed. Immediate advances would
include methods that take the dependence among chemical elements into account
within some generative model, or techniques that combine chemical abundances
with dynamical constraints to place joint prior probabilities on whether any
two stars could have formed from the same star cluster, given some model of the
Milky Way. 

In this work we focus on the former.
Here we present a new approach to chemical tagging that allows us to identify 
the latent (unobserved) factors that contribute to the chemical abundances of 
all stars (e.g., nucleosynthetic yields) while simultaneously performing 
clustering in the latent space. Notwithstanding caveats that we will
discuss in detail, this allows us to infer nucleosynthetic yields rather than
strictly prescribe them from models. Moreover, the scale of the clustering
problem reduces by a significant fraction because the clustering is performed in
a lower dimensional latent space instead of the higher dimensional data space.
In Section~\ref{sec:methods} we describe the model and the methods we use to
estimate the model parameters. Section~\ref{sec:experiments} describe the 
experiments performed using generated and real data sets. We discuss the results
of these experiments in Section~\ref{sec:discussion}, including the caveats with
the model as described. We conclude in Section~\ref{sec:conclusions}.

\section{Methods} \label{sec:methods}

Latent factor analysis is a common statistical approach for describing correlated 
observations with a lower number of latent variables \citep[e.g.,][]{Thompson:2004}.
Related techniques include principal component analysis \citep{Hotelling:1933} and its
variants \citep{Tipping;Bishop:1999}, singular value decomposition \citep{Golub:1970}, and other
matrix factorization methods. While factor analysis on its own is a useful
dimensionality reduction tool to identify latent factors that contribute to
the chemical abundances of stars \citep[e.g.,][]{Ting:2012,Price-Jones:2018,Milosavljevic:2018}, factor
analysis cannot describe clustering in the data (or latent) space. As a result,
some works have performed clustering and then required different latent factors for each (totally assigned) component \citep[e.g.,][]{EdwardsDowe1998}.
Similarly, clustering techniques applied to chemical abundances to date 
\citep[e.g.,][]{Hogg:2016} do not account for the lower effective dimensionality in
elemental abundances.

Here we expand on a variant of factor analysis known elsewhere as a mixture of common 
factor analyzers \citep{Baek:2010}, where the data are generated by a set of 
latent factors that are common to all data, but the scoring (or extent) of those
factors is different for each data point, and the data can be modelled as a
mixture of multivariate normal distributions in the latent space (factor scores).
In this work the data $\vecdataunscaled$ is a 
$\NumDimensions \times \NumData$ matrix where $\NumData$ is the number of 
stars and $\NumDimensions$ is the number of chemical abundances measured 
for each star. We assume a generative model for the data 
\begin{equation}
	\vecdataunscaled = \vec\mu + \factorloads\factorscores + \vec{e}
	\label{eq:generative-model}
\end{equation}

\begin{figure*}
	\includegraphics[width=\textwidth]{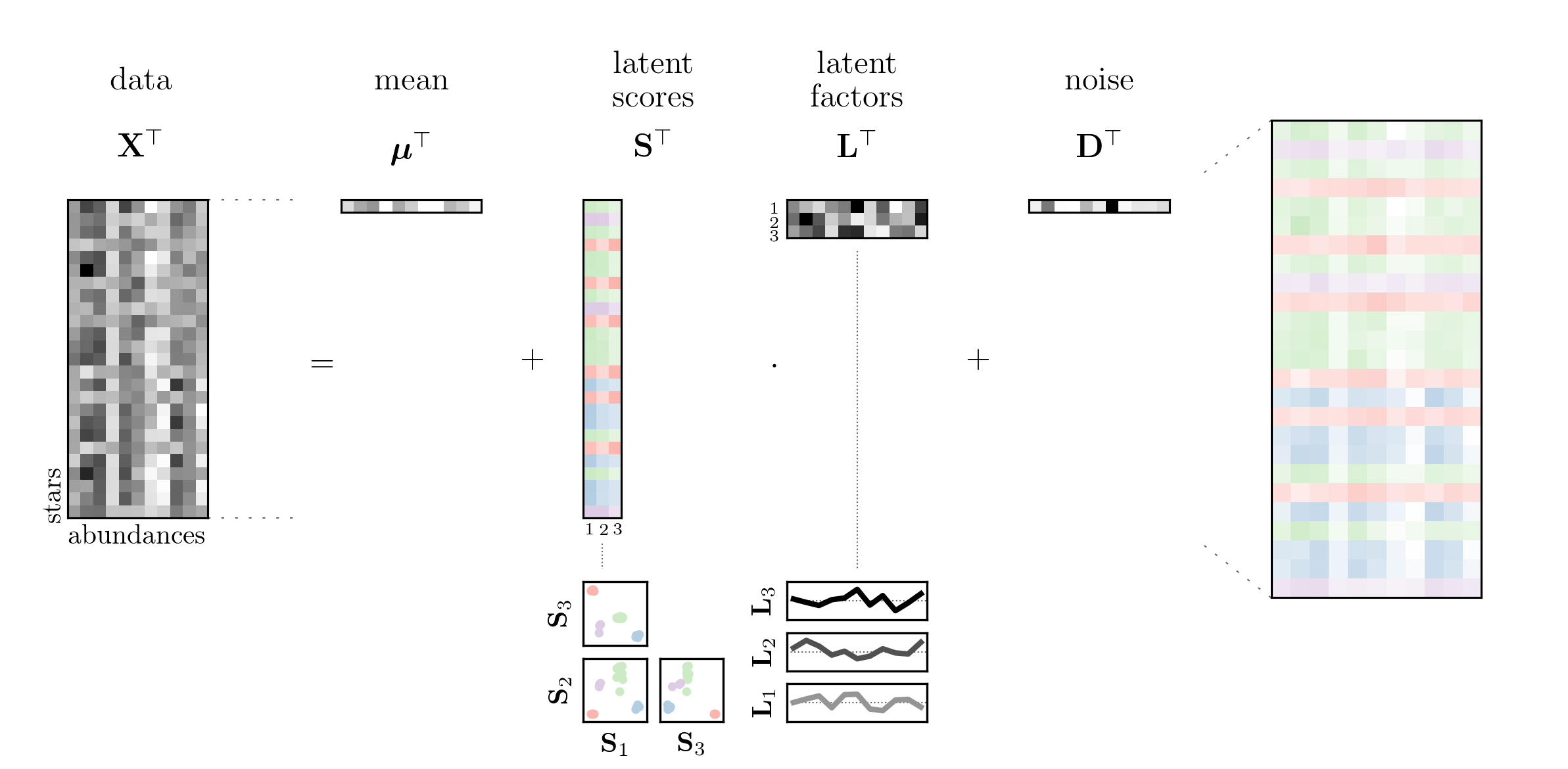}
	\caption{\added{A schematic that visualises the model components.
			 The data are shown on the left in greyscale, and magnified on the
			 right where each star (row) is coloured by its identified
			 component in latent space.
			 For each chemical abundance (column) there is a mean value and
			 variance that is independent of the latent space.
			 The latent factors $\factorloads$ are analogous to nucleosynthetic
			 yields and are common to all stars. The factor scores $\factorscores$
			 have an entry for each yield, for each star (row). The latent scores are modelled
			 by a mix of multivariate normal distributions of $\NumComponents$
			 components, which are coloured accordingly. The matrices of factor 
			 scores $\factorscores$ and factor loads $\factorloads$ are 
			 visualised at the bottom. For clarity here we show transposed
			 matrices (e.g., see Eq. \ref{eq:generative-model}).}}
	\label{fig:schematic}
\end{figure*}

\noindent{}where $\factorloads$ is a $\NumDimensions \times \NumLatentFactors$ 
matrix of factor loads that is common to all data points, $\NumLatentFactors$ is
the number of latent factors, and the factor scores 
for the $\numdata^\mathrm{th}$ data point
\begin{equation}
	\factorscores_\numdata \sim \mathcal{N}(\vec\xi_\numcomponents, \vec\Omega_\numcomponents)
\end{equation}
\noindent{}are drawn from\footnote{For clarifying nomenclature across disciplines, the terminology $z \sim \mathcal{N}(0, 1)$ indicates that the $z$ variable \emph{is drawn from} a standard normal distribution.} the $\numcomponents^\mathrm{th}$ multivariate
normal distribution, where each $\numcomponents^\mathrm{th}$ component has a $\NumLatentFactors$-dimensional mean and a dense $\NumLatentFactors \times \NumLatentFactors$ covariance matrix.
The mean vector $\vec\mu$ describes the mean datum in each dimension.
The factor scores for all data points $\factorscores$ is then a 
$\NumLatentFactors \times \NumData$ matrix, where each data point has a partial
association to the components in latent space. 
We assume $\vec{e} \sim \mathcal{N}\left(\vec{0}, \textrm{diag}(\specificvariance)\right)$
is independent of the latent space, and $\specificvariance$ is a
vector of variances in each $\NumDimensions$ abundance dimensions.
In this model each data point can be represented as being drawn
from a mixture of multivariate normal components, except the components
are \emph{clustered in the latent space} $\factorscores$ and projected
into the data space by the factor loads $\factorloads$. In a sense we
are using latent factor analysis as a form of dimensionality reduction and
simultaneously performing clustering in latent space \added{(Figure~\ref{fig:schematic})}.

We assume that the latent space is lower dimensionality than the
data space (i.e., $\NumLatentFactors < \NumDimensions$).
Within the context of stellar abundances, the factor loads
$\factorloads$ can be thought of as the \emph{mean} yields
of nucleosynthetic
events (e.g., $s$-process production from AGB stars averaged over
initial mass function and star formation history), and the
factor scores are analogous to the relative counts of those 
nucleosynthetic events. The clustering in factor scores
achieves the same as a clustering procedure in data space,
except we simultaneously estimate the latent processes that are
common to all stars (the so-called factor loads, analogous to 
nucleosynthetic yields). Within this framework a rare nucleosynthetic event
can still be described as a `factor load' $\factorloads_\numlatentfactors$, 
but its rarity would be represented by associated factor
scores being zero for most stars and thus have no contribution
to the observed abundances. In practice the factor loads can only be 
identified up to orthogonality and cannot be expressly interpreted as
nucleosynthetic yields because they have limited physical meaning
(we discuss this further in Section~\ref{sec:discussion}),
but this description of typical yields and relative event rates should
help build intuition for the model parameters, and provide context
within the astrophysical problem it is being applied.

Including latent factors in the model description allows us to account for 
processes that affect multiple elemental abundances. In this way we are 
accounting for the fact that the data dimensions are not independent of
each other. Another benefit is the scaling with computational cost. If we 
considered data sets of order $10^{7.5}$
entries (e.g., 30 chemical abundances for $10^6$ stars) purely as a
clustering problem, then even the most efficient clustering
algorithms would incur a significant cumulative computational 
overhead by searching the parameter space for the number of
clusters, and the optimal model parameters given that number
of components. However, because the mixture of factor analyzers
approach assumes that there is a \emph{lower dimensional latent 
space} in which the data are clustered, and that clustering is 
projected into real space by common factor loads, the 
dimensionality of the clustering problem is reduced from 
$N \times D$ to $N \times J$. This reduces computational cost through
faster execution of each optimization step, and on average fewer optimization steps
needed to reach a specified convergence threshold.

From a statistical standpoint, the primary advantage to using
a mixture of factor analysers is that we can simultaneously
estimate latent factors (e.g., infer nucleosynthetic 
yields) and perform clustering (e.g., chemical tagging) 
within a statistically consistent framework. That is to say
that we have a generative data-driven model that can 
quantitatively describe nucleosynthetic yields, and the
factor scores can explain the variance in turbulence and gas mixing,
or star formation efficiency, and the parameters of this model
can be simultaneously estimated in a self-consistent way with
a single scalar-justified objective function.

Without loss of generality the density of the mean-subtracted 
data $\vecdataunscaled - \vec\mu$ (which we hereafter will refer to simply as $\vecdata$) can be described as 
\begin{equation}
	f(\vecdata; \vec\Psi) = \sum_{\numcomponents=1}^{\NumComponents}\weight_\numcomponents\phi(\vecdata;\factorloads\scoremeans_\numcomponents, \factorloads\scorecovs_\numcomponents\factorloads\transpose + \diag{\specificvariance})
\end{equation}
\noindent{}given $\NumLatentFactors$ common factor loadings and $\NumComponents$ components
clustered in the latent (factor score) space. Here the parameter
vector
$\vec\Psi$ includes $\{\factorloads,\vec\pi,\scoremeans,\scorecovs,\specificvariance\}$, and $\phi(\vecdata; \vec\theta)$
describes the density of a multivariate Gaussian distribution,
and $\weight_\numcomponents$ describes the relative weighting of the $\numcomponents^\mathrm{th}$
component in latent space and $\sum\vec\weight_{\numcomponents} = 1$.
The log likelihood is then given by
\begin{equation}
	\log\mathcal{L}(\vecdata|\vec\Psi) = \sum_{\numdata=1}^{\NumData}\log{f(\vecdata;\vec\Psi)} \quad . \label{eq:log-likelihood}
\end{equation}

The model as described is indeterminate in that there is no unique 
solution for the factor loads $\factorloads$ and scores
$\factorscores$. These quantities can only be determined up until 
orthogonality in $\factorloads$. However, as we will describe in Section \ref{sec:expectation-maximization}, with suitable priors on $\vec\Psi$ 
one can efficiently estimate the model parameters using the expectation-maximization
algorithm \citep{Dempster:1977}. 

\vspace{1em}
\subsection{Initialisation} \label{sec:initialisation}

Here we describe how the model parameters are initialised.\footnote{This describes the default initialisation approach. Other approaches are available in the accompanying software.}
To initialise the factor loads $\factorloads$ we start by randomly drawing a $\NumDimensions \times \NumDimensions$ matrix from a Haar distribution \citep{Haar:1933},
which is uniform on the special orthogonal group $\textrm{SO}(n)$ and therefore guaranteed to return an orthogonal
matrix with a determinant of unity \citep{Stewart:1980}.
We denote the $\NumLatentFactors \times \NumDimensions$ left-most region\footnote{The region choice is arbitrary. All that is required is that the randomly-generated matrix have mutually orthogonal vectors.}
 of this
matrix to be our initial guess of $\factorloads$, which provides a set of mutually
orthogonal vectors.

We then initially assign each data point as belonging to one of the
$\NumComponents$ components using the \texttt{k-means++} algorithm \citep{Arthur:2007}
in the latent space. Given the initial
factor loads and assignments, we then estimate the relative weights
$\vec\pi$, the mean factor scores of each component $\scoremeans$, and
the covariance matrix of factor scores of each component $\scorecovs$.
Finally, we initialise the specific variance $\specificvariance$ in each
dimension as the variance in each data dimension. Other initialisation 
methods for the latent factors include singular value decomposition \citep{Golub:1970}
or generating random noise with orthogonal constraints, and random assignment
is an alternative method that is available for initialising assignments.

Throughout this work we repeat this initialisation procedure 25 times for
every trial of $\NumLatentFactors$ and $\NumComponents$ for a given data set. 
We then run expectation-maximization (Section~\ref{sec:expectation-maximization})
from each initialisation until the log likelihood improves by less than $10^{-5}$
per step, and we adopt the model with the highest log likelihood as the preferred 
model given that trial of $\NumLatentFactors$, $\NumComponents$, and the data. Although
this optimisation procedure is not convex, in practice it is normally sufficient to
initialise from many points to avoid local minima.

\subsection{Expectation-Maximization} \label{sec:expectation-maximization}

We use the expectation-maximization algorithm to estimate the model parameters
\citep{Dempster:1977}. With each expectation step we evaluate the log likelihood 
given the model parameters\footnote{When evaluating the log likelihood we use the precision (sparse inverse) matrix of the Cholesky decomposition of the covariance matrix for computational efficiency and stability.} $\vec\Psi$, the message length, and the $\NumData \times \NumComponents$ responsibility 
matrix $\vec\tau$ whose entries are the posterior probability that the 
$\numdata$th data point is associated to the $\numcomponents$th component, given 
the data $\vecdata$ and the current estimate of the parameter vector $\vec\Psi$:
\begin{equation}
	\tau_{\numdata\numcomponents} = \frac{\weight_\numcomponents\phi(\vecdata_\numdata;\factorloads\scoremeans_\numcomponents, \factorloads\scorecovs_\numcomponents\factorloads\transpose + \diag{\specificvariance})}{\sum_{g=1}^{G}\weight_g\phi(\vecdata_\numdata;\factorloads\scoremeans_g, \factorloads\scorecovs_g\factorloads\transpose + \diag{\specificvariance})} \quad .
\end{equation}

At the maximization step we update our estimates of the parameters $\vec\Psi$,
conditioned on the data $\vecdata$ and the responsibility matrix $\vec\tau$.
The updated parameters estimates are found by setting the second derivative
of the log likelihood (Eq.~\ref{eq:log-likelihood}) to zero and solving for
the parameter values.\footnote{Strictly this introduces a statistical inconsistency in that we should update our parameter estimates by setting the second derivative of our information-theoretic objective function (Eq.~\ref{eq:message-length}) to zero instead of the log likelihood, but this inconsistency only becomes serious with small $N$ \added{(e.g., $\approx 30$)} -- precisely the opposite situation of chemical tagging!}
In doing so this guarantees that every updated
estimate of the model parameters is guaranteed to increase the log likelihood.
Although there are no guarantees against converging on local minima, in 
practice it is sufficient to run expectation-maximization from multiple
initialisations (as we do) in order to ensure that the global minimum is reached.
At the maximization step we first update our estimate of the relative weights 
$\vec\weight\nextstep$ given the responsibility matrix $\vec\tau$
\begin{equation}
	\weight_\numcomponents\nextstep = \frac{1}{\NumData} \sum_{\numdata=1}^{\NumData}\tau_{\numdata\numcomponents}
\end{equation}
\noindent{}where the $\vec{\Psi}\thisstep$ superscript refers to the current parameter estimates and $\vec{\Psi}\nextstep$ refers to the updated estimate for the next iteration.
The updated estimates of the mean factor scores 
$\scoremeans\nextstep$ for each component are then given by
\begin{eqnarray}
	\scoremeans_\numcomponents\nextstep = \scoremeans_\numcomponents\thisstep + \frac{\vec{G}\transpose(\vecdata\transpose - \factorloads\thisstep\scoremeans_\numcomponents\thisstep)\vec\tau_\numcomponents}{\NumData\weight_\numcomponents\nextstep}
\end{eqnarray}
\noindent{}where:
\begin{eqnarray}
	\vec{W} &=& (\scorecovs_\numcomponents\thisstep)^{-1} \\
	\vec{V} &=& \left(\specificvariance\thisstep\right)^{-1} \\
	\vec{C} &=& (\vec{W} + (\factorloads\thisstep)\transpose\vec{V}\factorloads\thisstep)^{-1} \\
	\vec{G} &=& \left[\vec{V} - \vec{V}\factorloads\thisstep\vec{C}\left(\vec{V}\factorloads\thisstep\right)\transpose\right]\factorloads\thisstep\scorecovs_k\thisstep \quad .
\end{eqnarray}

The covariance matrices of the components of factor scores $\scorecovs\nextstep$
are updated next,
\begin{equation}
	\scorecovs_\numcomponents\nextstep = \left(\eye - \vec{G}\transpose\factorloads\thisstep\right)\scorecovs_\numcomponents\thisstep + \frac{\vec{G}\transpose\vec{Z}\left(\vec{Z}\vec\tau_\numcomponents\transpose\right)\transpose\vec{G}}{N\weight_\numcomponents\nextstep}
\end{equation}
\noindent{}where
\begin{eqnarray}
	\vec{Z} &=& \vecdata\transpose - \factorloads\thisstep\scoremeans_\numcomponents\thisstep \quad .
\end{eqnarray}

After some linear algebra, updated estimates of the common factor loads $\factorloads\nextstep$
can be found from
\begin{equation}
	\factorloads\nextstep = \factorloads_{a}\left(\factorloads_{b}^{-1}\eye\right)
\end{equation}
\noindent{}where:
\begin{eqnarray}
	\factorloads_\textrm{a} &=& \sum_{\numcomponents=1}^{\NumComponents}\left[ \vec\tau_\numcomponents\transpose\vecdata\left(\scoremeans_\numcomponents\thisstep\right)\transpose + 
	\vec{G}\transpose\vec\tau_\numcomponents\vec{Z}\transpose\vec{G}\right] \\
	\factorloads_\textrm{b} &=& N\sum_{\numcomponents=1}^{\NumComponents}\left[\weight_\numcomponents\nextstep\left(\scorecovs_\numcomponents\nextstep + \scoremeans_\numcomponents\nextstep\left(\scoremeans_\numcomponents\nextstep\right)\transpose\right)\right] \quad
\end{eqnarray}

Finally, the updated estimate of the specific variances $\specificvariance\nextstep$ are given
by
\begin{equation}
	\specificvariance\nextstep = \frac{1}{\NumData}\left[\sum^{\NumComponents}_{\numcomponents=1}\vec\tau_\numcomponents\transpose\left(\vecdata\odot\vecdata\right) - \sum_{j=1}^{J}\left(\factorloads\nextstep\factorloads_\textrm{b}\right)\odot\factorloads\nextstep\right]
\end{equation}

\noindent{}where $\odot$ denotes the entry-wise product. 
Throughout this work we assume that the data are noiseless and
we do not add any observed errors to the constructed covariance matrices.
% We comment on the effects of this assumption in Section~\ref{sec:discussion}.

\subsection{Missing data}

The expectation-maximization procedure as described requires that there be no 
missing data entries in order to update our estimates of the responsibility matrix $\vec\tau$
and the model parameters $\vec\Psi$.
In practice, however, there will often be abundance measurements that are missing
for some subset of stars. There are many potential reasons for this, including
astrophysical explanations (e.g., an absorption line was not present above the noise),
observational limitations (e.g., the signal-to-noise ratio was too low, or 
contamination by a cosmic ray), or various other reasons that cannot be inferred
from the available information.

In this work we will assume that any missing data measurements are
missing at random. The missing data points can then be treated as unknown
parameters that must be solved for (and updated) at each iteration. Initially
we impute zeros for missing data entries in $\vec\data$, and at each iteration we update
these imputed value with our estimate of what the missing data values
are given the current model parameters. This ensures that the log likelihood increases
with each iteration. Similarly, with each update we inflate
our estimates of the specific variances based on the fraction of missing
data points in each dimension
\begin{equation}
\specificvariance_d\nextstep = \specificvariance_d\nextstep\left(\frac{N}{N-\replaced{M}{M_d}}\right)
\end{equation}
\noindent{}where \replaced{$M$}{$M_d$} is a the number of missing data entries in the $d$th dimension.
In Section \ref{sec:toy-model-missing-data} we show with a toy model that the latent
factor loads and scores can be reliably estimated even in the presence of high fractions
of missing data (e.g., 40\%), conditioned on our assumption that the data are missing
at random.

\subsection{Model Selection}

The expectation-maximization algorithm as described requires a specified 
number of latent factors $\NumLatentFactors$ and $\NumComponents$. In the next 
Section we describe a toy model using generated data where we will assume
that the true number of latent factors and components are not known. 
We require some heuristic to decide how many latent factors and components are 
preferred given some data. An increasing number of factors
and components will undoubtedly increase the log likelihood of the model given
the data, but the log likelihood does not account for the increased model 
complexity that is afforded by those additional latent factors and components.

One criterion commonly employed for evaluating a class of models is the 
Bayesian Information Criterion \citep[BIC;][]{Schwarz:1978}
\begin{equation}
	\textrm{BIC} = Q\log{N} - 2\log\mathcal{L}\left(\data|\vec\Psi\right)
	 \label{eq:bic}
\end{equation} 
\noindent{}where $Q$ is the number of parameters in this model
\begin{equation}
	Q = \frac{\NumLatentFactors}{2}\left[2\left(\NumDimensions - \NumLatentFactors\right) + \NumComponents\left(3 + \NumLatentFactors\right)\right] + \NumComponents + \NumDimensions - 1
\end{equation}
\noindent{}which includes $\NumComponents - 1$ weights (as $\sum\weight_\numcomponents = 1$), $\NumDimensions$ specific variances, the $\NumComponents\NumLatentFactors - \NumLatentFactors^2$ free parameters needed to uniquely define the mutually orthogonal factor loads matrix $\factorloads$ \citep{Baek:2010}, $\NumComponents \times \NumLatentFactors$ parameters for the mean scores $\scoremeans$, and $\frac{1}{2}\NumComponents\NumLatentFactors\left(\NumLatentFactors + 1\right)$ parameters to encode the $\NumComponents$ full $\NumLatentFactors \times \NumLatentFactors$ covariance matrices $\scorecovs$.

While the BIC does include a penalisation term for the number of
parameters (which scales with $\log{N}$), it does not describe for the
increased flexibility that is afforded by the addition of those parameters.
For example, adding one parameter
to a model will increase the BIC by at most $\log{N}$, but there are different
ways for a single parameter to be introduced. In a fictitious model $y=f(x)$
a parameter $b$ could be added that is a scalar multiple of $x$, or it could be
introduced as $x^b$. Despite the difference in model complexity, the same
penalisation occurs in BIC. Even if the log likelihood were only to improve
marginally in both cases, the difference in model complexity is not captured
by BIC. In other words, there are situations where we are more interested in
balancing the model complexity (or the expected Fisher information and similar
properties) with the goodness of fit, instead of penalising the number of 
parameters.

For these reasons we use the Minimum Message Length \citep[MML;][]{Wallace:2005}
principle as a criterion for model selection and evaluation. 
The classically-described principle of MML is that the best explanation of the
data given a model is the one that leads to the shortest so-called two-part message~\citep{Wallace:2005}, 
where a \textit{message} takes into account both the complexity of the model 
and its explanatory power. The complexity of the model is described through
the first part of the message, and the second part of the message describes
its explanatory power. The \emph{length} of each message part is quantified
(or estimated) using information theory, allowing for a fair evaluation between
different models of varying complexity or explanatory power. MML has been 
shown to perform well on a variety of empirical analyses 
\citep[see, e.g.,][]{WallaceDowe1994b,viswanathan1999finding,EdwardsDowe1998,WallaceDowe2000,fitzgibbon2004minimum,Wallace:2005,dowe2007bayes,Dowe2008a,Dowe2011a}.
Arguments about the statistical consistency (i.e.,~as the number of data 
points increases the distributions of the estimates become increasingly 
concentrated near the true value) of MML are given in \citet{DoweWallace1997a,Dowe2011a}.
The MML principle requires that we explicitly specify our prior beliefs on the
model parameters, providing a Bayesian optimisation approach which can be
applied across entire classes of models.

The \textit{message} must encode two parts: the model, and the data given the
model. The encoding of the message is based on Shannon's information theory~\citep{Shannon:1948}. 
The information gained from an event $e$ occurring, where $p(e)$ is the
probability of that event, is $I(e) = -\log_{2}{p(e)}$. The information content
is largest for improbable outcomes, and smallest for outcomes that we are 
almost certain about. In other words, an outcome that has a probability close
to unity has nearly zero information content because almost nothing new is learned from it,
whereas rarer events convey a much higher information content.

In practice calculating the message length can be a non-trivial task, 
especially for models that are reasonably complex. This can make the strict MML
principle intractable in many cases and necessitates
approximations to the message length
\citep[however see][]{Wallace:1987,WallaceDowe1999a,Wallace:2005}. Using a Taylor expansion, a generalised
scheme can be calculated to estimate the parameter vector $\vec\Psi$ that
minimises the message length ${I}(\vec\Psi,\vecdata)$ \citep{Wallace:1987},
\begin{equation}
	{I}(\vec\Psi,\vecdata) = \frac{Q}{2}\log\kappa_Q - \log\left(\frac{p(\vec\Psi)}{\sqrt{|\mathcal{F}(\vec\Psi)|}}\right) - \log\mathcal{L}\left(\vecdata|\vec\Psi\right) + \frac{Q}{2} \label{eq:mml} 
\end{equation}
\noindent{}where $\log\likelihood(\vecdata|\vec\Psi)$ is the familiar
log likelihood, $p(\vec\Psi)$ is the joint prior density on $\vec\Psi$,
$\mathcal{F}(\vec\Psi)$ is the matrix whose entries are the expected second order
partial derivatives of the log likelihood,
commonly referred to as the expected Fisher information matrix,
\begin{equation}
	\mathcal{F}(\vec\Psi) = -\textrm{E}\left[\frac{\partial^2}{\partial\vec\Psi^2}\log\likelihood(\vecdata|\vec\Psi)\right]
\end{equation}
\noindent{}and as before $Q$ is the number of model parameters.
Continuous parameters can only be stated to finite precision, which leads
to the $\frac{Q}{2}\log\kappa_Q$ term that gives a measure of the volume of the region of
uncertainty in which the parameters $\vec\Psi$ are centred. The $\log\kappa_Q$
term can be reasonably approximated by
\begin{equation}
	\log\kappa_Q = -\log{2\pi} + \frac{1}{Q}\log{Q\pi} - \gamma - 1
\end{equation}
\noindent{}where $\gamma$ is Euler's constant.

Like the BIC, the message length is penalised by the number of model parameters
through the $\log\kappa_Q$ term. However, the model complexity is also 
described through the priors and the Fisher information matrix, which
describes the curvature of the log likelihood with respect to the model
parameters. For these reasons, MML provides a more accurate description of
the model complexity (or flexibility) because it naturally includes the 
curvature of the log likelihood with respect to the model parameters
rather than only penalising models based on the \emph{number} of
parameters.

We will describe the contributions to the message length in parts. We assume
the priors on the number of latent factors $\NumLatentFactors$ and the
number of components $\NumComponents$ to be
$p(\NumLatentFactors) \propto 2^{-\NumLatentFactors}$ and
$p(\NumComponents) \propto 2^{-\NumComponents}$ respectively,
such that fewer numbers are preferred. The optimal lossless message to 
encode each is \citep[Sec. 6.8.2;][]{Knorr-Held:2000},
\begin{eqnarray}
	I(\NumLatentFactors) &= -\log{p(\NumLatentFactors)} &= \NumLatentFactors\log{2} + \textrm{constant} \label{eq:prior_J} \\
	I(\NumComponents) &= -\log{p(\NumComponents)} &= \NumComponents\log{2} + \textrm{constant} \label{eq:prior_K} \quad .
\end{eqnarray}
Only $\NumComponents - 1$ of the relative weights $\vec\weight$ need encoding because 
$\sum_{\numcomponents=1}^{\NumComponents}\weight_\numcomponents = 1$. We assume a uniform prior on 
individual weights,
\begin{equation}
	p(\vec\weight) = (\NumComponents - 1)!
\end{equation}
\noindent{}and the Fisher information is
\begin{equation}
	\mathcal{F}(\vec\weight) = \frac{\NumData^{\NumComponents - 1}}{\prod_{\numcomponents=1}^{\NumComponents}\weight_\numcomponents}
\end{equation}
\noindent{}which gives the message length of the relative weights $I(\vec\weight)$ to be
\begin{eqnarray}
	I(\vec\weight) &=& -\log\left(\frac{p(\vec\weight)}{\sqrt{|\mathcal{F}(\vec\weight)|}}\right) \nonumber \\
			   &=& -\log{p(\vec\weight)} - \frac{1}{2}\log{|\mathcal{F}(\vec\weight)|} \nonumber  \\
			   &=& -\log(\NumComponents - 1)! + \frac{\NumComponents - 1}{2}\log{\NumData} - \frac{1}{2}\sum_{\numcomponents=1}^{\NumComponents}\log\weight_\numcomponents \nonumber \\
	I(\vec\weight) &=& \frac{1}{2}\left(\left(\NumComponents - 1\right)\log\NumData - \sum_{\numcomponents=1}^{\NumComponents}\log\weight_\numcomponents\right) -\log\Gamma(\NumComponents) \quad . \nonumber \label{eq:prior_pi} \\
\end{eqnarray}

We assume uniform priors for the component means in latent space $\scoremeans$, 
where the bounds are large enough outside the range of observable values such
that those priors are proper (integrable) -- a necessary condition for the MML principle -- 
and only add constant terms to the message length, which can be ignored. We assume 
a conjugate inverted Wishart prior for the component covariance matrices
$\scorecovs$ \citep{WallaceDowe1994b,WallaceDowe2000,Knorr-Held:2000},
\begin{equation}
	p(\scoremeans_\numcomponents,\scorecovs_\numcomponents) \propto |\scorecovs_\numcomponents|^{\frac{1}{2}(\NumLatentFactors + 1)} \quad .
\end{equation}

We approximate the determinate of the Fisher information of a multivariate normal $|\mathcal{F}(\scoremeans,\scorecovs)|$
as $|\mathcal{F}(\scoremeans)||\mathcal{F}(\scorecovs)|$ \citep{Oliver:1996,Figueiredo:2002} where
\begin{equation}
	|\mathcal{F}(\scoremeans)| = (\NumData\weight_k)^\NumLatentFactors|\scorecovs_k|^{-1}
\end{equation}
\begin{equation}
	|\mathcal{F}(\scorecovs)| = (\NumData\weight_k)^{\frac{1}{2}\NumLatentFactors(\NumLatentFactors+1)}2^{-\NumLatentFactors}|\scorecovs_k|^{-(\NumData+1)}
\end{equation}

\noindent{}such that 

\begin{eqnarray}
	I(\scoremeans,\scorecovs) &=& -\sum_{\numcomponents=1}^{\NumComponents}\log{p(\scoremeans_k,\scorecovs_k)} + \frac{1}{2}\sum_{\numcomponents=1}^{\NumComponents}\log{|\mathcal{F}(\scoremeans_k,\scorecovs_k)|} \nonumber \\
I(\scoremeans,\scorecovs) &=& \frac{1}{4}\NumLatentFactors(\NumLatentFactors+3)\sum_{\numcomponents=1}^\NumComponents\log{\NumData\weight_k} - \frac{KD}{2}\log{2} \nonumber \\ 
&& \cdots \,\, -\frac{1}{2}(2\NumLatentFactors+3)\sum_{k=1}^{K}\log{|\scorecovs_k|}  \quad . \label{eq:prior_xi_omega} 
\end{eqnarray}

Previous work on multiple latent factor analysis within the context of MML have
addressed the indeterminacy between the factor loads and factor scores by
placing a joint prior on the \emph{product} of factor loads and scores \citep{WallaceMLF,EdwardsDowe1998,Wallace:2005}.
Adopting the same prior density in our model is not practical because 
it would require the priors $p(\scoremeans|\vec\tau,\vec\weight$) and $p(\scorecovs|\vec\tau,\vec\weight$).
That is, we would require a prior density on both the means $\scoremeans$
and covariance matrices $\scorecovs$ in latent space that requires knowledge
about the responsibility matrix $\vec\tau$ and relative weights $\vec\weight$ in order to estimate the effective scores
$\factorscores$ for each data point and calculate a joint prior on the product
of the factor loads $\factorloads$ and factor scores $\factorscores$.
Instead we address this indeterminacy
by placing a prior on $\factorloads$ that ensures it is mutually orthogonal.
Specifically, we adopt a Wishart distribution with scale matrix $\vec{W}$
and $D$ degrees of freedom for the
$\NumLatentFactors\times\NumLatentFactors$ matrix $\vec{M} = \factorloads\transpose\factorloads$.
In other words, $\vec{M} \sim W_\NumLatentFactors(D,\vec{W})$
and $\vec{W} = \textrm{Cov}(\factorloads\transpose)$.
This
Wishart joint prior density gives highest support for mutually orthogonal vectors,
\begin{equation}
	p(\factorloads) = \frac{|\factorloads\transpose\factorloads|^{\frac{1}{2}(\NumDimensions - \NumLatentFactors - 1)}}{2^{\frac{\NumDimensions\NumLatentFactors}{2}}|\vec{W}|^{\frac{\NumDimensions}{2}}\Gamma(\frac{\NumDimensions}{2})}\exp\left[-\frac{1}{2}\textrm{Tr}(\vec{W}^{-1}\factorloads\transpose\factorloads)\right] \quad .
\end{equation}

\begin{widetext}
\noindent{}Thus the message length to encode $\factorloads$ is given by
\begin{eqnarray}
I(\factorloads)	&=& \frac{1}{2}\textrm{Tr}(\textrm{Cov}(\factorloads\transpose)^{-1}\factorloads\transpose\factorloads) - \frac{1}{2}(\NumDimensions - \NumLatentFactors - 1)\log{|\factorloads\transpose\factorloads|} + \frac{1}{2}\NumDimensions\NumLatentFactors\log{2} + \frac{1}{2}\NumDimensions\log{|\textrm{Cov(\factorloads)}|} - \Gamma\left(\frac{\NumDimensions}{2}\right) \quad . \label{eq:prior_L}
\end{eqnarray}
%\end{widetext}
Combining equations \ref{eq:prior_J}, \ref{eq:prior_K}, \ref{eq:prior_pi}, \ref{eq:prior_xi_omega}, and \ref{eq:prior_L} with equation \ref{eq:mml} leads to the full message length:
%\begin{widetext}
\begin{eqnarray}
	I(\vec\Psi, \vec\data) &=& -\log\likelihood(\vec\data|\vec\Psi)
 +\frac{1}{4}\left(\NumLatentFactors + 4\right)\left(\NumLatentFactors - 1\right)\sum_{\numcomponents=1}^\NumComponents\log\weight_\numcomponents + \left(\NumComponents - \frac{1}{2}\right)\log{\NumData}
 +\frac{1}{2}\NumDimensions\log|\textrm{Cov}\left(\factorloads\transpose\right)| \nonumber \\
  && \cdots -\frac{1}{2}\left(D-J-1\right)\log|\factorloads\transpose\factorloads| + \textrm{Tr}\left(\textrm{Cov}\left(\factorloads\transpose\right)^{-1}\factorloads\transpose\factorloads\right) - \left(\NumLatentFactors + \frac{3}{2}\right)\sum_{\numcomponents=1}^\NumComponents\log|\scorecovs_\numcomponents|  -\log\Gamma\left(\NumComponents\right) - \Gamma\left(\frac{\NumDimensions}{2}\right) \nonumber \\
&& \cdots +\frac{Q}{2}\log\kappa_q +\frac{1}{2}\left[\NumLatentFactors(\NumDimensions + 2) + \NumComponents(2-\NumData)\right]\log{2}  \quad . \label{eq:message-length}
 \end{eqnarray}
 \newpage
\end{widetext}

\section{Experiments} \label{sec:experiments}

\subsection{A toy model} \label{sec:exp-1}

Here we introduce a toy model where we use generated data to verify that
we recover the true model parameters given some data, and to
ensure that the expectation-maximization method is yielding consistent results.
We generated a data set with ${\NumData = 1}$00,000 data points, each with
$\NumDimensions = 15$ dimensions. We adopted a latent dimensional space of 
$\NumLatentFactors = 5$ factor loads such that the vector $\factorloads$ has
shape $\NumDimensions \times \NumLatentFactors$,
with $\NumComponents = 10$ clusters in the latent space. 
We generated the random factor loads in the same way that we initialise the optimisation (Section~\ref{sec:initialisation}). The relative weights $\vec\weight$
are drawn from a multinomial distribution and the means of the clusters
in factor scores $\scoremeans$ are drawn from a standard normal
distribution. The off-diagonal entries in the covariance matrices in factor scores $\scorecovs$ are drawn from a gamma distribution $\scorecovs_{\numcomponents,i,i} \sim \vec\Gamma\left(1\right)$. The variance in 
each dimension $\specificvariance$ are also drawn $\specificvariance \sim \vec\Gamma\left(1\right)$.
The $\numdata$th data point (which belongs to the $\numcomponents$th cluster) is then
generated by drawing $\factorscores_{\numdata} \sim \mathcal{N}(\scoremeans_\numcomponents,\scorecovs_\numcomponents)$, projecting by the factor loads $\factorloads$, and adding variance $\specificvariance$.

We treat the generated data set as if the number of latent factors
and components are not known. Starting with $\NumLatentFactors = 1$
and $\NumComponents = 1$, we trialled each permutation of $\NumLatentFactors$ and $\NumComponents$
until $\NumLatentFactors_\textrm{max} = 10$
and   $\NumComponents_\textrm{max} = 20$ (e.g., twice the true values of $\NumLatentFactors_\textrm{true}$ and $\NumComponents_\textrm{true}$).

\begin{figure*}
	\centering
	\begin{tabular}[b]{@{}p{0.45\textwidth}@{}}
		\centering\includegraphics[width=\linewidth]{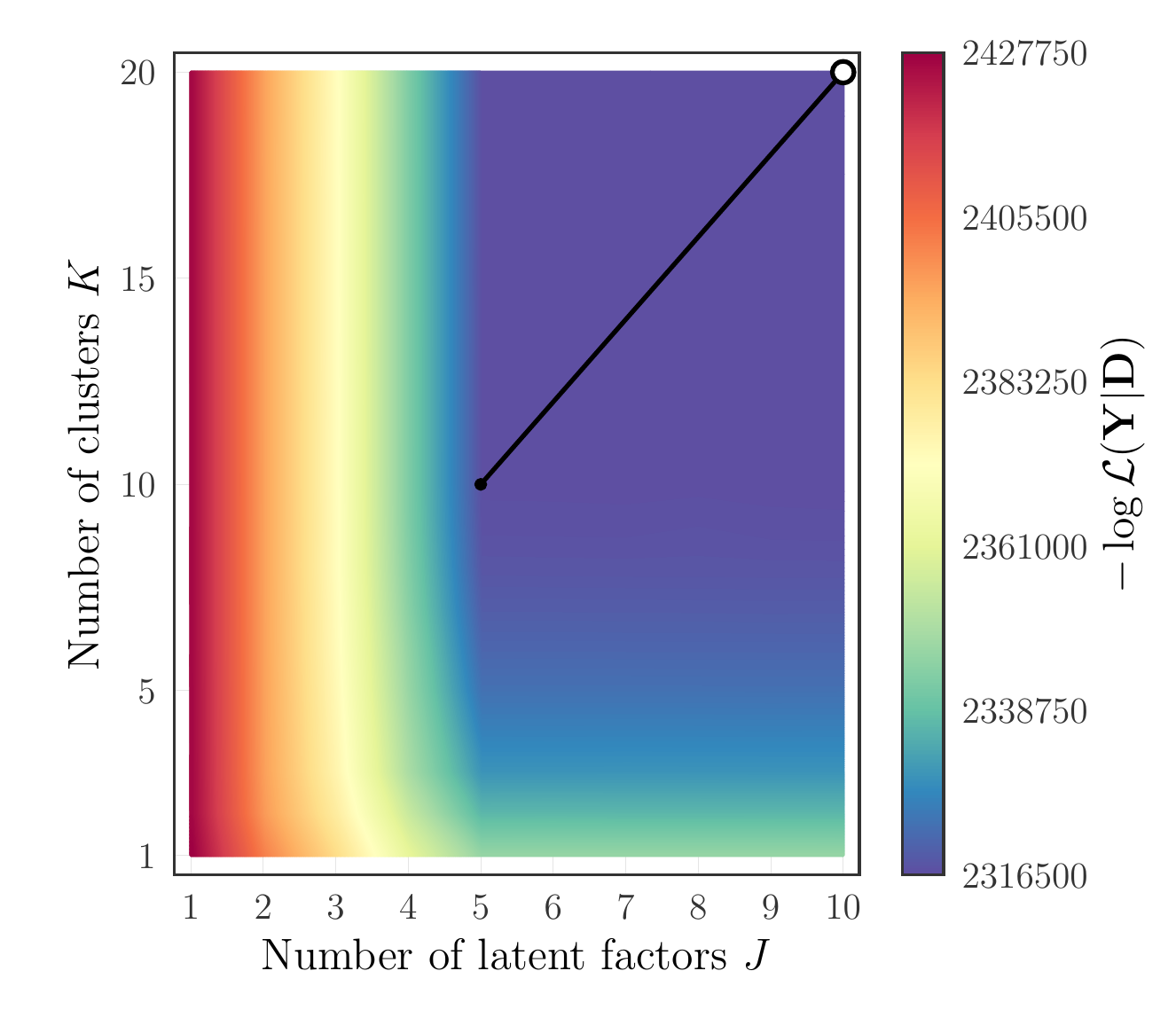} \\
	\end{tabular}
	\begin{tabular}[b]{@{}p{0.45\textwidth}@{}}
		\centering\includegraphics[width=\linewidth]{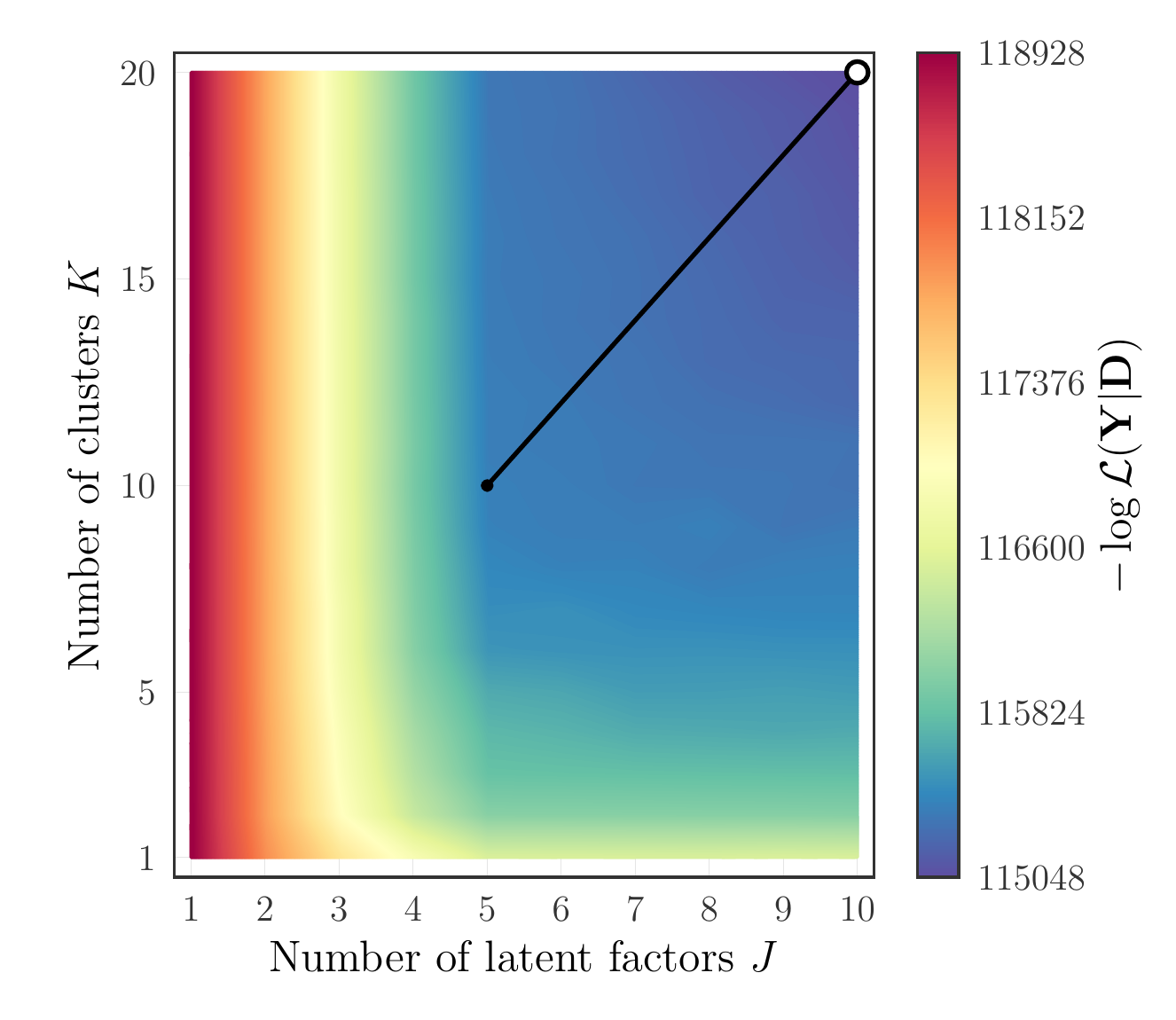} \\
	\end{tabular}

	\begin{tabular}[b]{@{}p{0.45\textwidth}@{}}
		\centering\includegraphics[width=\linewidth]{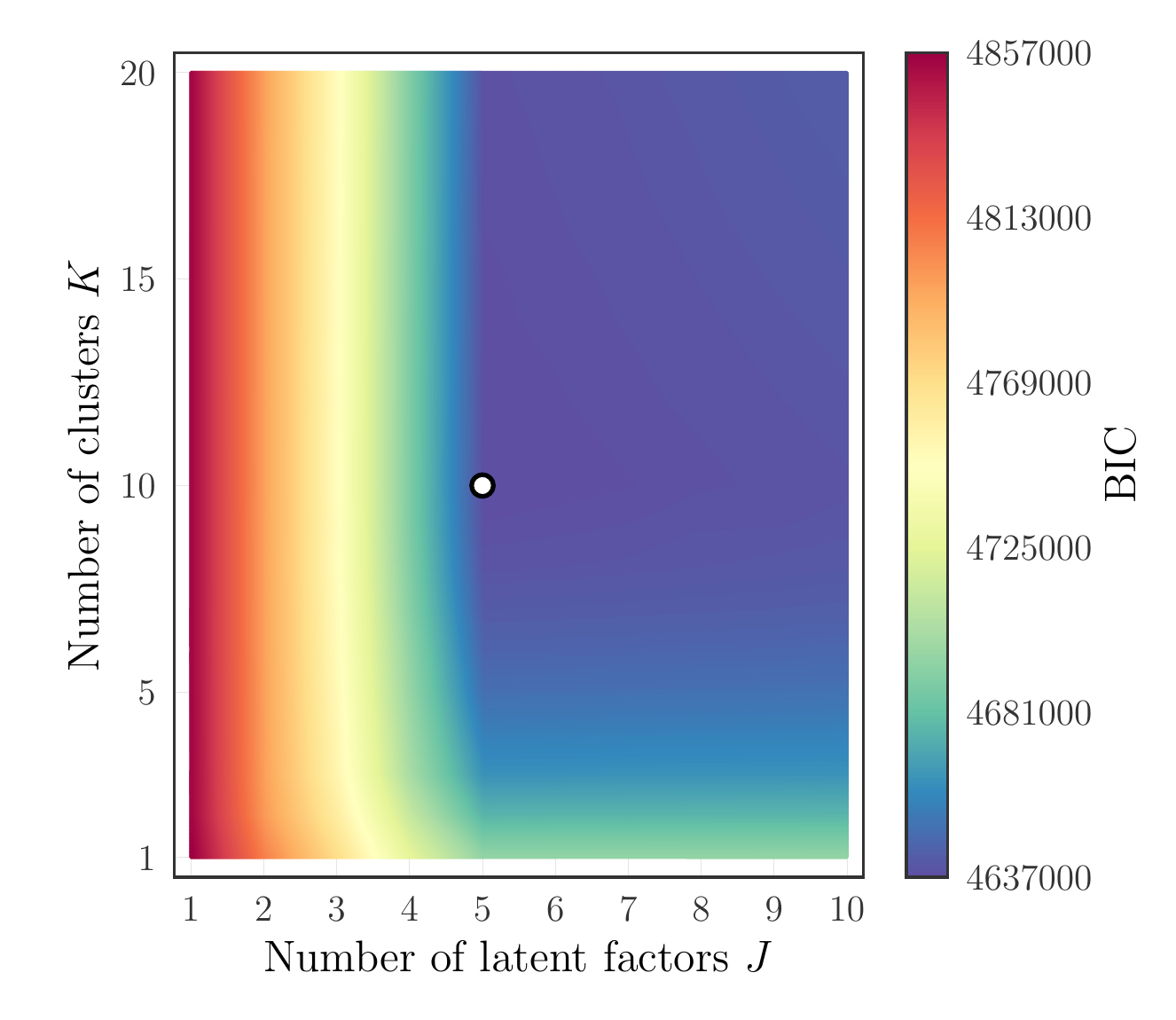} \\
	\end{tabular}
	\begin{tabular}[b]{@{}p{0.45\textwidth}@{}}
		\centering\includegraphics[width=\linewidth]{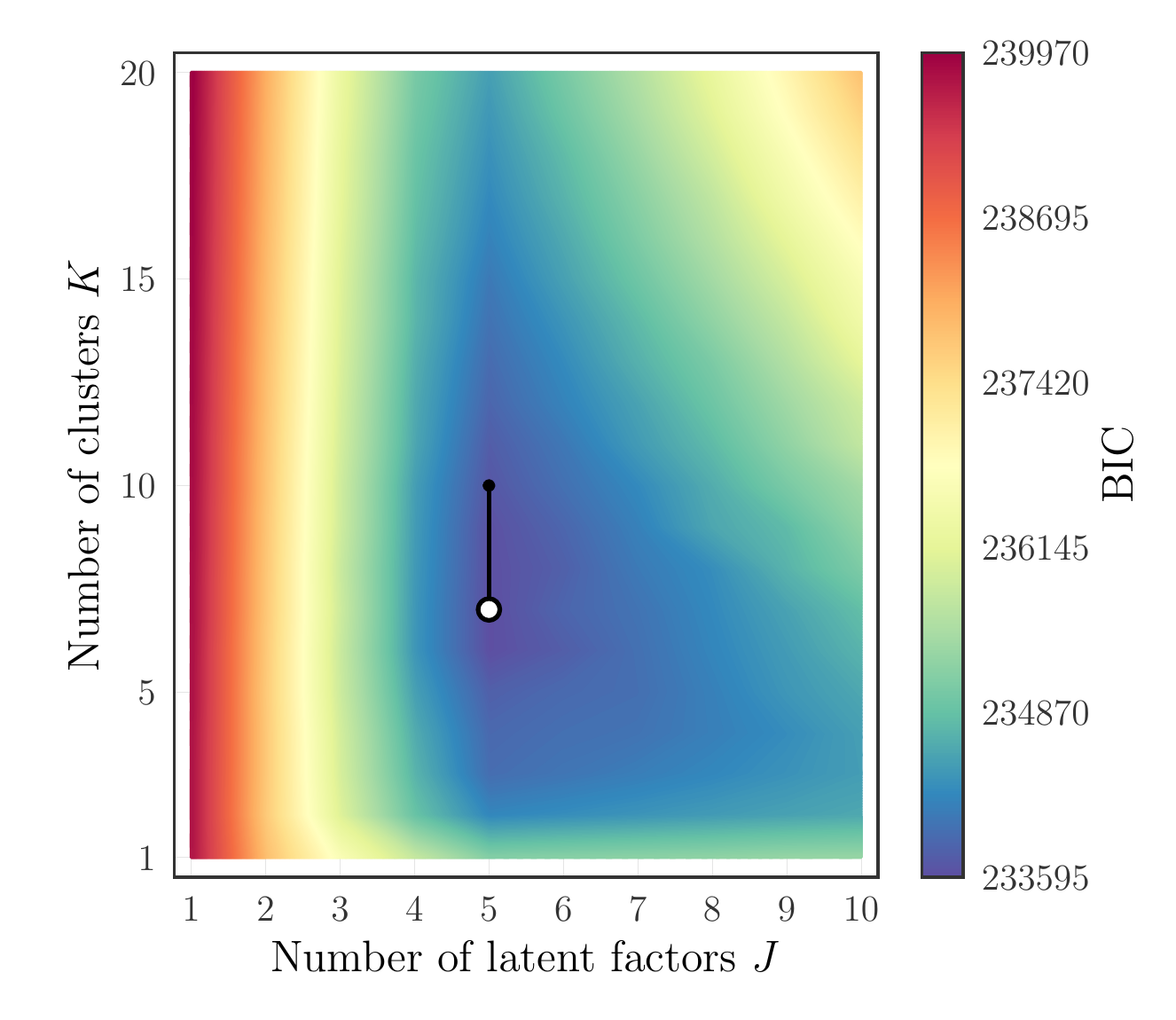} \\
	\end{tabular}

	\begin{tabular}[b]{@{}p{0.45\textwidth}@{}}
		\centering\includegraphics[width=\linewidth]{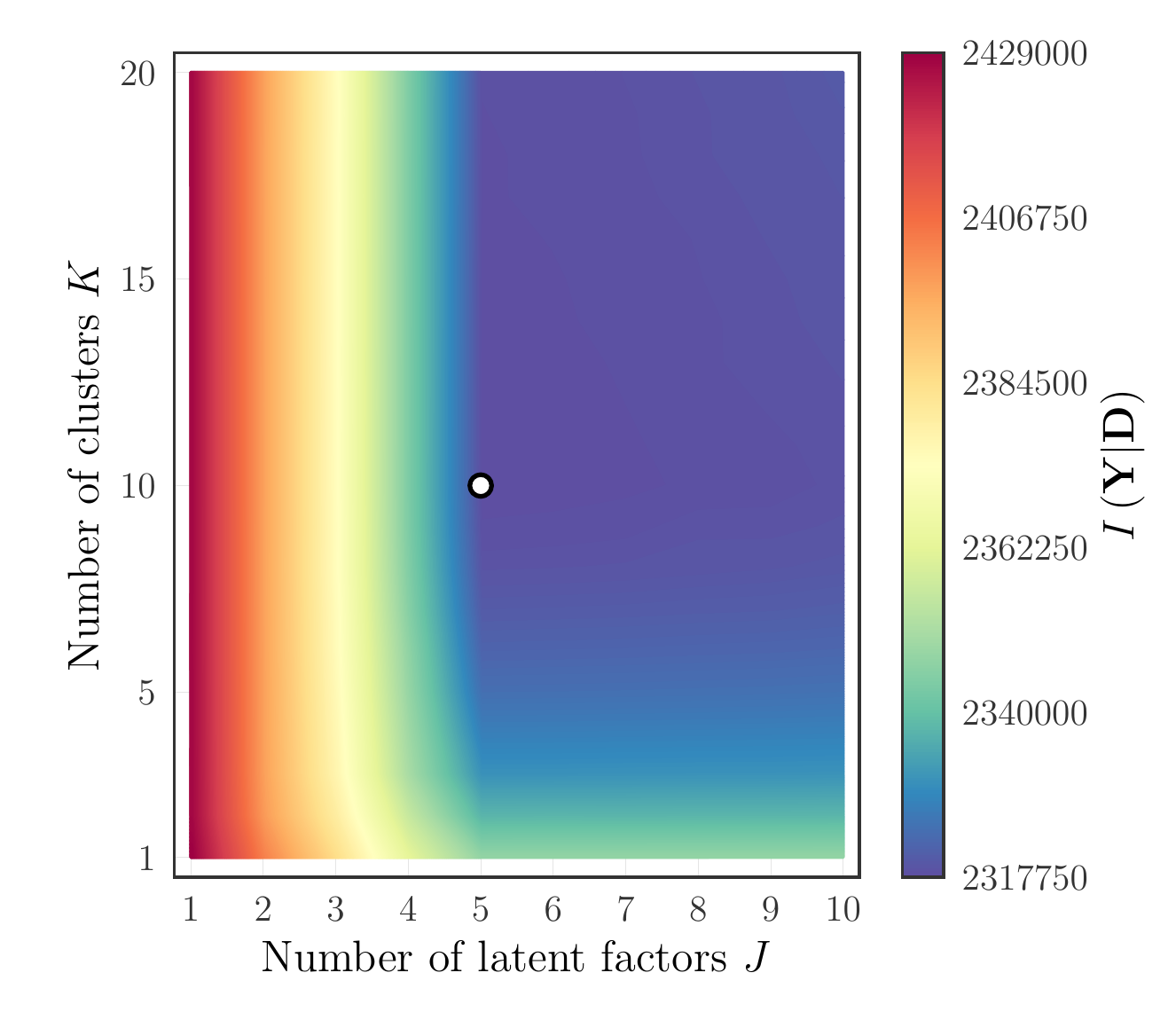} \\
	\end{tabular}
	\begin{tabular}[b]{@{}p{0.45\textwidth}@{}}
		\centering\includegraphics[width=\linewidth]{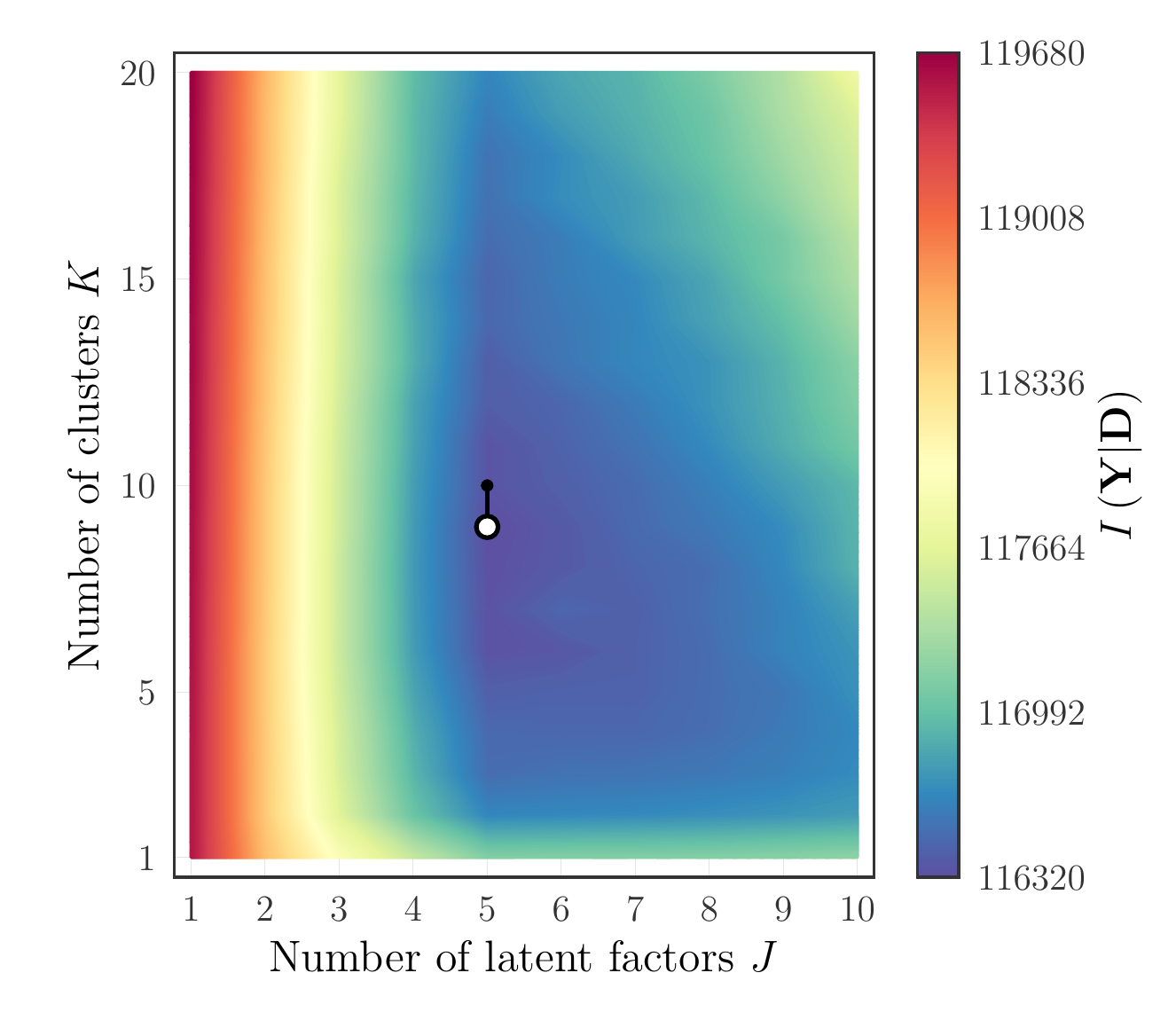} \\
	\end{tabular}
    \caption{Metrics from our grid search for the toy model for two sample sizes: 
    		 $N = 100,000$ (left) and $N = 5,000$ (right). 
    		 The top panels show the negative log likelihood
			 $-\log{\mathcal{L}\left(\data|\vec\Psi\right)}$ 
			 evaluated at each combination of latent factors 
			 $\NumLatentFactors$ and number of clusters 
			 $\NumComponents$ using the generated data in our toy model.
			 The middle panels shows 
			 the BIC (Eq.~\ref{eq:bic}) for those 
			 combinations, and the lower panel shows the 
			 message length. The white marker indicates the
			 lowest value in each panel. A line connects to the true value (black point) to guide the eye.}
    \label{fig:experiment-1-gridsearch}
    \vspace{-10em}
\end{figure*}

We recorded the \emph{negative} log likelihood, the BIC, and the message length\footnote{Omitting constant terms such that negative message lengths are allowed.}  for each permutation of $\NumLatentFactors$ and $\NumComponents$.
These metrics are shown in Figure~\ref{fig:experiment-1-gridsearch}.
Unsurprisingly the negative log likelihood decreases with increasing numbers of latent
factors $\NumLatentFactors$ and increasing numbers of components $\NumComponents$.
The lowest BIC value and message length is found at $\NumLatentFactors = 5$
and $\NumComponents = 10$, identical to the true values. 

\added{We repeated this toy model experiment using a smaller sample size ($N = 5,000$) to be more representative of the sample sizes in later \Galah\ experiments (Section~\ref{sec:exp4}). The results of the grid search are also shown in Figure~\ref{fig:experiment-1-gridsearch}. Here BIC estimates the true number of latent factors correctly, but tends to underestimate the true number of clusters, more so than the message length. Although the difference between the true number of components and that given by the shortest message length is not large, this does serve to illustrate that in this example a larger number of data points are required to `resolve' the true number of components in latent space.}

It is clear from Figure~\ref{fig:experiment-1-gridsearch} that a combination of latent factors
and clustering in the latent space provides a better description of the (generated) data than a Gaussian mixture model without latent factors.
Adding components to the model does improves the log likelihood, even with a single latent factor,
but the addition of just \emph{one latent factor} improves the log likelihood more so than adding
\emph{twenty components}. Not much more can be said for this example because the true data generating process 
is known, but this toy model does illustrate how 
clustering in high dimensional data can be better described by latent factors with 
clustering in the lower dimensional latent space.

Some technical background is warranted before we compare our estimated model
parameters to the true values. We previously stated that the latent factors in this model are only
identifiable up to an orthogonal rotation. That is to say that
if the data were truly generated by latent factors $\factorloads_\textrm{true}$,
then our estimates of those latent factors $\factorloads_\textrm{est}$ do not need
to be identical to the true values. For example, the ordering of the estimated factors
could be different from the true factors, and the ordering of the dimensionality
in latent space would then be accordingly different. Since no constraint is
placed on the ordering of the factor loads during expectation-maximization,
there is no assurance (or requirement) that our factor loads match the true factor loads.

Another possibility is that the estimated factor loads could be flipped in sign 
relative to the true factor loads, and the scores would similarly be flipped. 
In both of these situations (reordering or flipped signs) the log likelihood 
given the data and the estimated factor loads $\factorloads_\textrm{est}$ 
would be identical to the log likelihood given the data and the true factor loads 
$\factorloads_\textrm{true}$
despite the difference in ordering and sign. The same can be said for any other
scalar metric \citep[e.g., Kullback-Leibler divergence;][]{Kullback:1951}.
These examples serve to illustrate a more 
general property that the factor loads and factor scores can be orthogonally 
rotated by \emph{any valid rotation matrix}\footnote{Recall that a rotation matrix is valid if 
$\vec{R}\vec{R}\transpose = \vec{I}\,\,$.} $\vec{R}$. The estimated factor loads 
$\factorloads_\textrm{est}$ could therefore appear very different from the true 
values, but they only differ by an orthogonal rotation. We discuss the impact of this limitation on real data in more detail in Section~\ref{sec:discussion}.

\begin{figure*}[t!]
	\includegraphics[width=\textwidth]{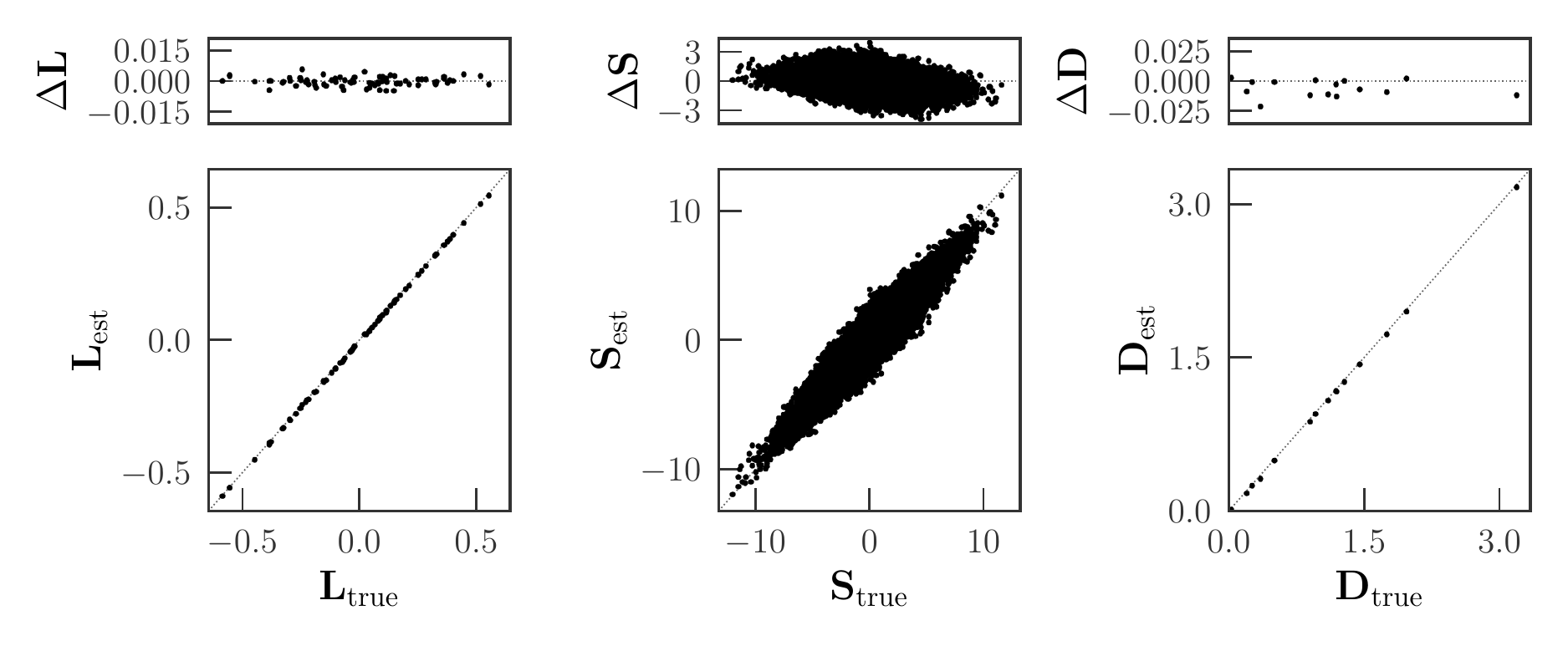}
	\caption{The estimated factor loads $\factorloads$ (left), factor scores $\factorscores$ (middle),
    		 and specific variances $\specificvariance$ (right) compared to the 
		 	 true data generating values
		 	 for Experiment~1 (Section~\ref{sec:exp-1}). The agreement is excellent.}
    \label{fig:exp1-compare}
\end{figure*}

We took the model with the preferred number of latent factors and components found
from a grid search ($\NumComponents = 10$, $\NumLatentFactors = 5$; which are also
the true values) and applied an orthogonal rotation to the latent space to be as
close as possible to the true values. The rotation matrix $\mathbf{R}$ was found
by solving for $\NumLatentFactors$ unknown angle parameters, each of which is used
to construct a Givens rotation matrix \citep{Givens:1958}, and then we take the product of those Givens
matrices to produce a valid rotation matrix $\vec{R}$. This process reduces to Euler angle rotation in three or fewer dimensions.
This process rotates the latent space
($\factorloads$, $\scoremeans$, $\scorecovs$), but has no effect on the model's 
predictive power: the evaluated log likelihood or the Kullback-Leibler divergence \citep{Kullback:1951} under the
rotated model is indistinguishable from the unrotated model.
In Figure~\ref{fig:exp1-compare} we show the estimated factor loads $\factorloads$,
factor scores $\factorscores$, and specific variances $\specificvariance$ compared
to the true values. The agreement is excellent in all model parameters.

\subsection{A toy model with data missing at random}
\label{sec:toy-model-missing-data}

Here we repeat the toy model used in the previous experiment, but we discard an increasing fraction
of the data and evaluate the performance and accuracy of our method in the presence
of incomplete data. We considered missing data fractions from 1\% to
40\%. In each case we treated the model parameters as unknown, assumed
the missing data points were missing at random, and initialised the
model as per Section~\ref{sec:exp-1}.

\begin{figure*}[t!]
	\includegraphics[width=\textwidth]{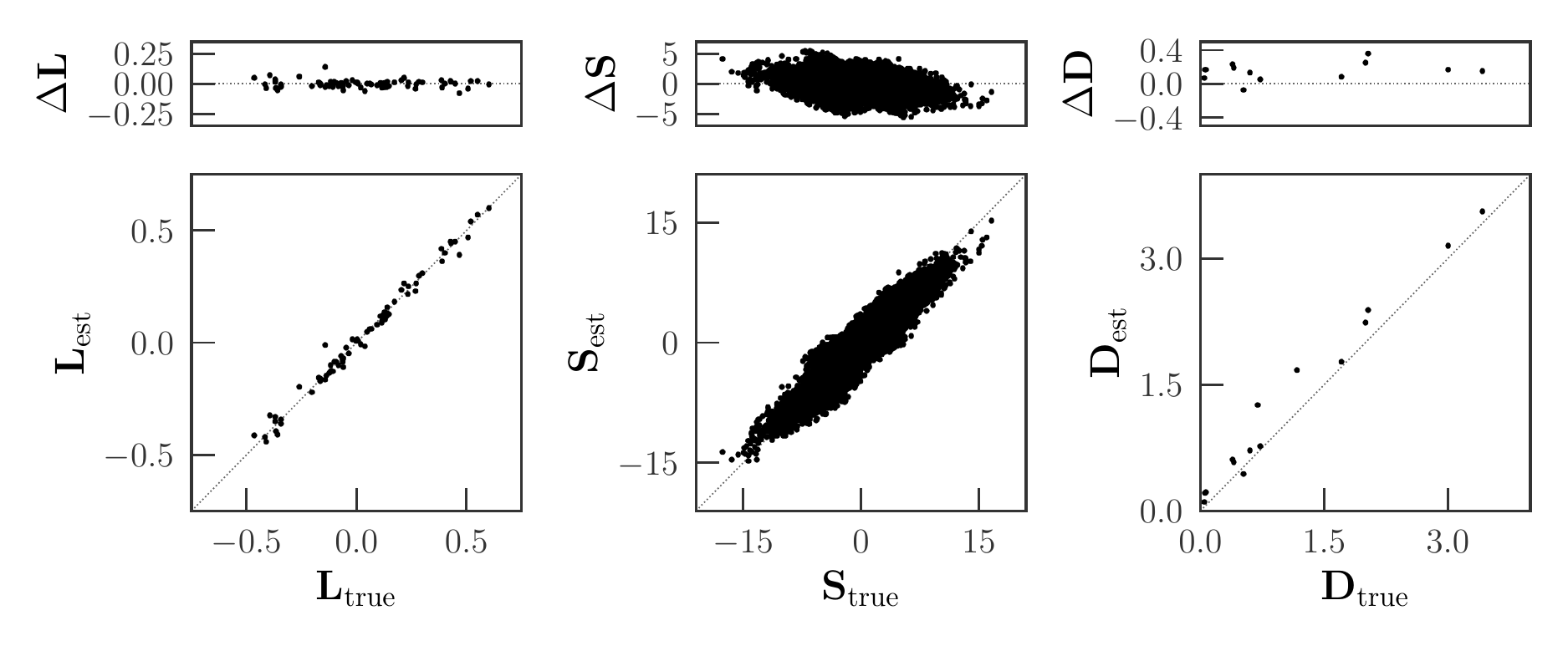}
	\caption{The estimated factor loads $\factorloads$ (left), factor scores $\factorscores$ (middle),
    		 and specific variances $\specificvariance$ (right) compared to the 
		 	 true data generating values
		 	 for Experiment~2 (Section~\ref{sec:toy-model-missing-data}). Here 40\% of the data are missing at random.
			 The agreement remains excellent,
			 despite the large fraction of missing data entries. Note that the scales on the top panels are 2-10 times larger than those in Figure~\ref{fig:exp1-compare}.}
    \label{fig:exp1-missing-data}
\end{figure*}

In Figure~\ref{fig:exp1-missing-data} we show the results of this
experiment for our worst considered case, where 40\% of the data
entries are randomly discarded. We find that despite the high fraction
of missing entries, our estimates of the model parameters remain unbiased
in this example using a toy model. The corrections to our estimates of the
specific variances are sufficient, in that the specific variance in each
dimension is not systematically under-estimated from the true values, 
despite that 40\% of the data entries are missing.

\subsection{The \Galah\ survey}
\label{sec:exp4}

In this experiment we perform blind chemical tagging using the 
photospheric abundances released as part of the second \Galah\ 
data release \citep{Buder:2018}. This data set includes
up to 23 chemical abundances reported for 342,682 stars.
In this example we chose to restrict ourselves to stars with a
complete set of abundance measurements for a subset of those 23 elements
(i.e., no missing data entries).
For example, here we will exclude
lithium and carbon abundances because the
photospheric values will vary throughout a star's lifetime \citep[e.g.,][]{Casey:2016b,Casey:2019}. This is true
to a small degree for many elements \citep[e.g.,][]{Dotter:2017},
but for the purposes of this experiment we assume that all other
photospheric abundances remain constant throughout a star's
lifetime.

We first selected stars with \texttt{flag\_cannon = 0} to exclude
stars where there is reason to suspect that the stellar parameters
(e.g., $\teff$, $\logg$) are unreliable, and as a result the 
detailed chemical abundances would be untrustworthy. 
We then required all stars to have a signal-to-noise ratio exceeding 
30 per pixel in the blue arm (\texttt{snr\_c1 > \replaced{40}{30}}). \added{This is equivalent
to a signal-to-noise ratio of about 140 per resolution element in
the third \project{HERMES} CCD ($\lambda_\textrm{central} \approx 5750\,$\AA).}
We required that stars have no erroneous flags in all of the following abundances: 
Mg, Na, Al, Si, K, Ca, Sc, Ti, Mn, Fe, Ni, Cu, Zn, Y, Ba, La, and Eu.
These elements were chosen because they trace multiple nucleosynthetic pathways, and they are more commonly reported in the \Galah\ data release, allowing for a larger number of stars with a complete abundance inventory.
There are \replaced{1,072}{2,566} stars that met these criteria. 

\added{We note that while our signal-to-noise ratio cut is arbitrary, it is in
part motivated by the point where systematic uncertainties start to
dominate in \Galah\ results \citep[Figure 15 of][]{Buder:2018}. Systematic
uncertainties per abundance can be captured by the specific variances $\specificvariance$
in our model. A more restrictive signal-to-noise cut would reduce the sample size and restrict our ability to infer latent factors and components, whereas a more relaxed signal-to-noise cut would still require
that there are no erroneous flags in abundances. }

\begin{figure}
	\includegraphics[width=0.45\textwidth]{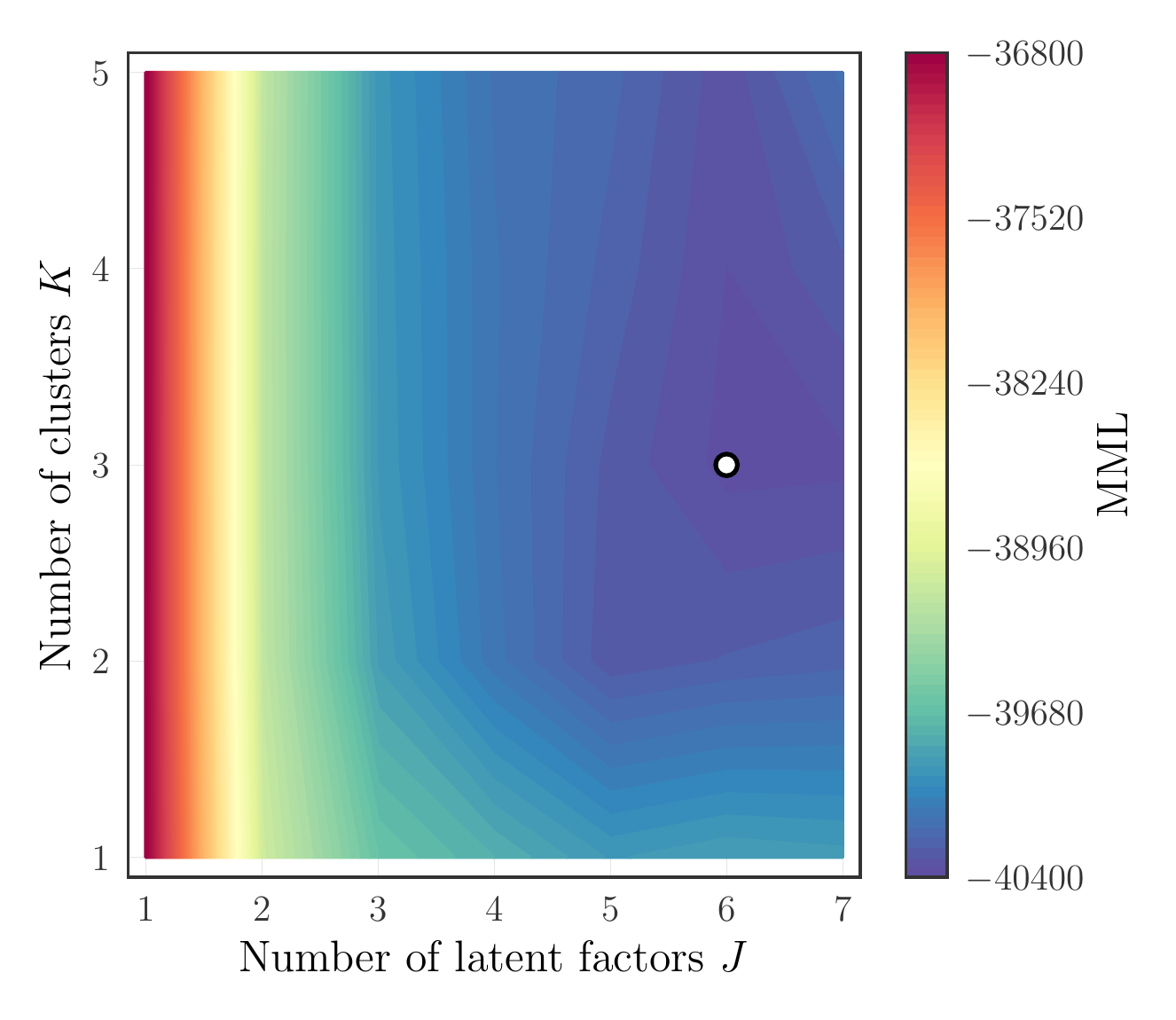}
    \caption{The top panel shows the negative log likelihood 
			 $-\log{\mathcal{L}\left(\data|\vec\Psi\right)}$ 
			 evaluated at each combination of latent factors 
			 $\NumLatentFactors$ and number of clusters 
			 $\NumComponents$ using \Galah\ data in
			 Experiment~3.  The middle panel shows 
			 the BIC for those combinations, and the lower panel shows the 
			 message length. The white marker indicates the
			 lowest value in each panel, showing the
			 preferred number of latent factors and components.}
    \label{fig:exp3-gridsearch}
\end{figure}

We executed a grid search for the number of latent factors $\NumLatentFactors$
and the number of components $\NumComponents$ that were preferred by the data.
Starting with $\NumLatentFactors = 1$ and $\NumComponents = 1$, we trialled each
permutation of $\NumLatentFactors$ and $\NumComponents$ up until $\NumLatentFactors = 7$
and $\NumComponents = 5$. 
The results of this grid search are shown in Figure~\ref{fig:exp3-gridsearch},
where we show the negative log likelihood, the BIC, and message length found for each permutation.
The behaviour of the BIC and the message length are very different here, unlike what was observed
in our toy model. Here the BIC behaviour appears similar to the negative log likelihood in that the BIC
prefers higher components and latent factors than the extent of the grid (e.g., $J > 7$ and $K > 5$). Indeed, if we were to trial higher values of $J$ and $K$ then the negative log likelihood would continue to increase.
The model with \replaced{five}{six} latent factors and three components (\replaced{$J=5$}{$J = 6$}, $K = 3$) is found to have the shortest message length, which we take as our preferred model for these data.

Earlier we described how the latent factors we estimate can only be identified up until an orthogonal rotation. If we want to interpret the latent factors estimated from \Galah\ data, then we must specify
some target factor loads such that we can identify which factors are most similar to the yields we expect.
We specified the following target latent factors where:
\begin{itemize}
	\item The first factor load should have non-zero entries in Eu and La (e.g., the $r$-process).
	\item The second factor load should have non-zero entries in Ba, Y, and La (e.g., the $s$-process).
	\item The third factor load should have non-zero entries in Fe-peak elements \deleted{V,} Mn, Fe, Ni, \replaced{and Zn}{Zn, and Ti}.
	\item The fourth factor load should have non-zero entries in the odd-Z elements Na, Al, K\added{, Sc, and Cu}.
	\item The fifth factor load should have non-zero entries in the $\alpha$-element tracers Si, Ca, and \replaced{Ti}{Mg}.
\end{itemize}

\begin{figure*}
	\includegraphics[width=\textwidth]{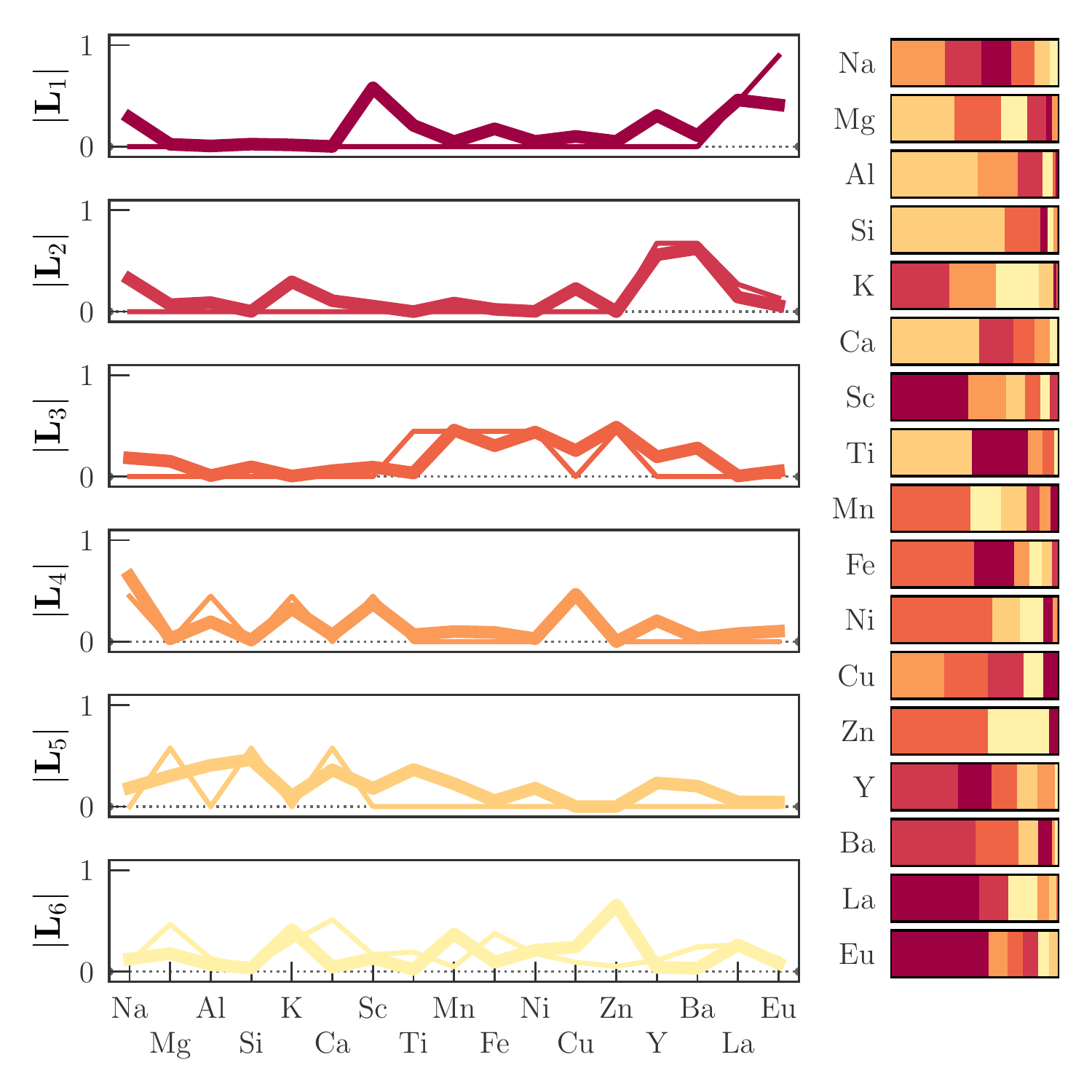}
	\caption{Latent factors inferred from 2,566 stars in \Galah\
			 \citep[][thick lines]{Buder:2018} with 17 abundance measurements. Left panels show the absolute entries for each
			 factor load, where the thin lines indicate the target latent factors (see Section~\ref{sec:exp4}). On the right we show the absolute fractional contributions
			 to each element, ordered by the loads that contribute most.}
    \label{fig:exp3-factor-loads}
\end{figure*}

We initially set each non-zero entry in these target factor loads $\factorloads_\textrm{target}$ to $E^{-\frac{1}{2}}$, where $E$ is the number of non-zero entries in that factor load, to ensure that $\factorloads_\textrm{target}$ is mutually orthogonal.
We solved for the $\NumLatentFactors$ unknown angles to produce a valid rotation matrix $\vec{R}$ that would make our estimated loads $\factorloads$ as close as possible to the target loads $\factorloads_\textrm{target}$, and then applied that rotation to the model. The target loads and (rotated) estimated loads are shown in Figure~\ref{fig:exp3-factor-loads}. Note that the purpose of this procedure is not to `find' the target loads that we expect, but to provide as little information needed in order to identify and describe all factor loads within an astrophysical context. This procedure still requires that the factors be mutually orthogonal and that they describe the data. For these reasons, we will not always recover the exact target loads we seek: we will only be able to identify factor loads that are closest to the target loads.

This is demonstrated in Figure~\ref{fig:exp3-factor-loads}, where some estimated factor loads match
closely to the target load (e.g., $\mathbf{L}_2$ which we identify as the s-process), and some barely
match at all (e.g., $\mathbf{L}_5$). Here we show the absolute entry of the factor loads because even
if an entry is negative, the corresponding factor scores could also be negative, and their product will contribute to the observed abundances. For this reason the sign does not matter here.

\begin{figure}
	\includegraphics[width=0.45\textwidth]{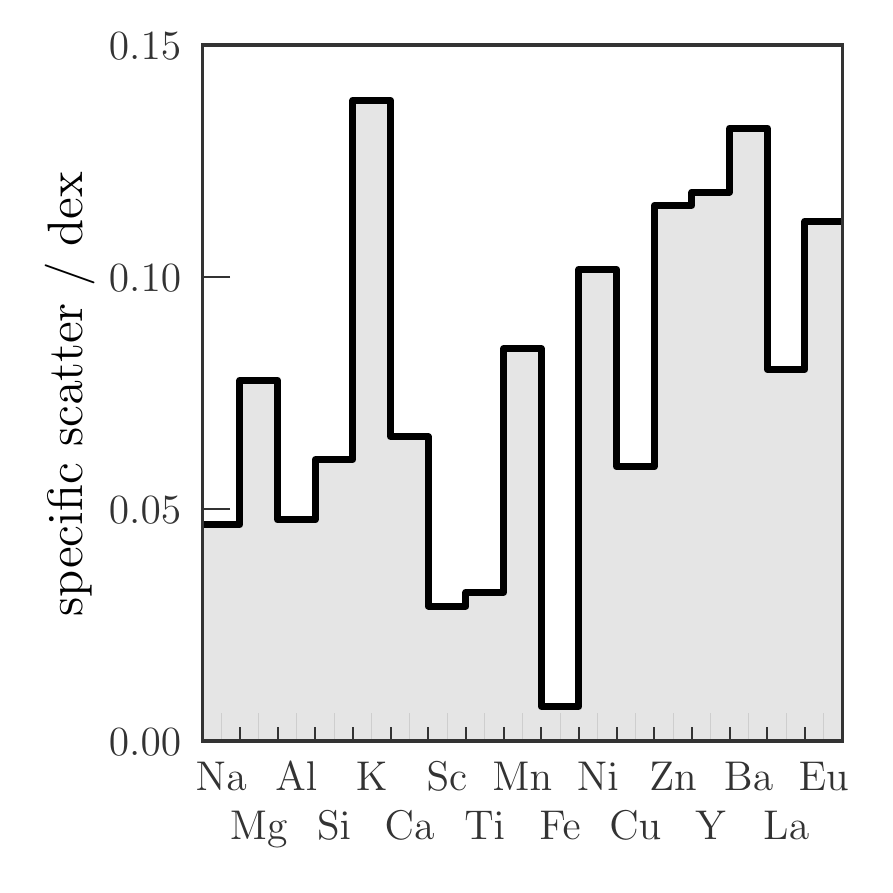}
	\caption{Specific scatter (e.g., $\sqrt{\specificvariance}$) remaining in the \Galah\ data \citep{Buder:2018}
			 after accounting for the contributions by all
			 latent factors.}
    \label{fig:exp3-specific-scatter}
\end{figure}

Some of these factor load entries may be non-zero because we require the latent factors to
be mutually orthogonal, and not because they truly contribute to the data. To try and
disentangle these possibilities, we calculate the fractional contribution that factor
load makes to the observed abundances relative to other factor loads. We define the
fractional contribution of the $j$th factor load to the $d$th
data dimension as:
\begin{equation}
	\mathbf{C}_\textrm{d,j} = \frac{\sum^{N}|\factorloads_{j,d}\factorscores_{n,j}|}{\sum^J\sum^{N}|\factorloads_{j,d}\factorscores_{n,j}|} \quad .
\end{equation}
The fractional contributions to each element are shown in the right hand side of
Figure~\ref{fig:exp3-factor-loads}. We identify the first factor $\mathbf{L}_1$ 
as being most similar to the r-process, and here it is the dominant contributor to Eu, a typical r-process tracer. Surprisingly we also find that this factor load is a reasonable contributor to the \replaced{Fe-peak}{odd-Z} element Sc. The specific scatter in Sc is 0.03~dex (Figure~\ref{fig:exp3-specific-scatter}), suggesting that the Sc abundances are well-described by this latent factor model.

The second latent factor $\mathbf{L}_2$ here is most representative of the slow neutron capture process (s-process), with dominant contributions to Ba, and Y. This factor has some support at other elements, notably K. $\mathbf{L}_3$ is the primary contributor to nearly all Fe-peak elements, with close to negligible contributions from other factors. The exception here is Cu, where a near-equal contribution comes from $\mathbf{L}_4$.
The fifth latent factor $\mathbf{L}_5$ is the dominant contributor to the $\alpha$-element tracers Si, Ca, and Mg, and surprisingly, Al. 
The specific scatter after accounting for these latent factors is smallest
for Fe (0.01~dex) and largest for K (0.13~dex; Figure~\ref{fig:exp3-specific-scatter}). The typical scatter in most elements is about 0.05~dex.

\begin{figure*}
	\includegraphics[width=\textwidth]{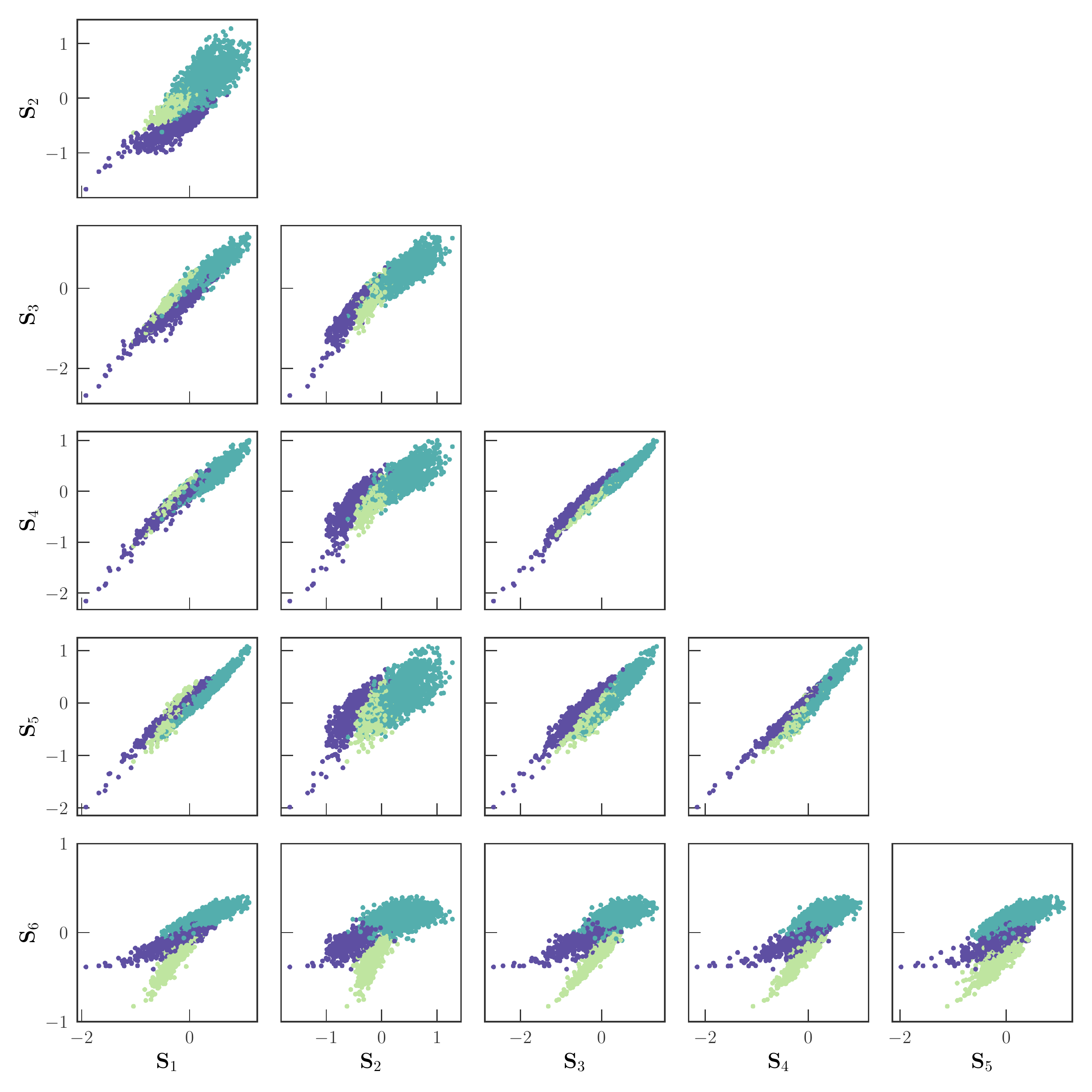}
	\caption{The factor scores $\factorscores$ estimated in Experiment~3 (Section~\ref{sec:exp4} using $N =$ \replaced{1,072}{2,566} stars in the \Galah\ data \citep{Buder:2018} that have 17 abundance measurements. Here each star is coloured by its inferred $K$th component.}
    \label{fig:exp3-latent-space}
\end{figure*}

\begin{figure*}
	\includegraphics[width=\textwidth]{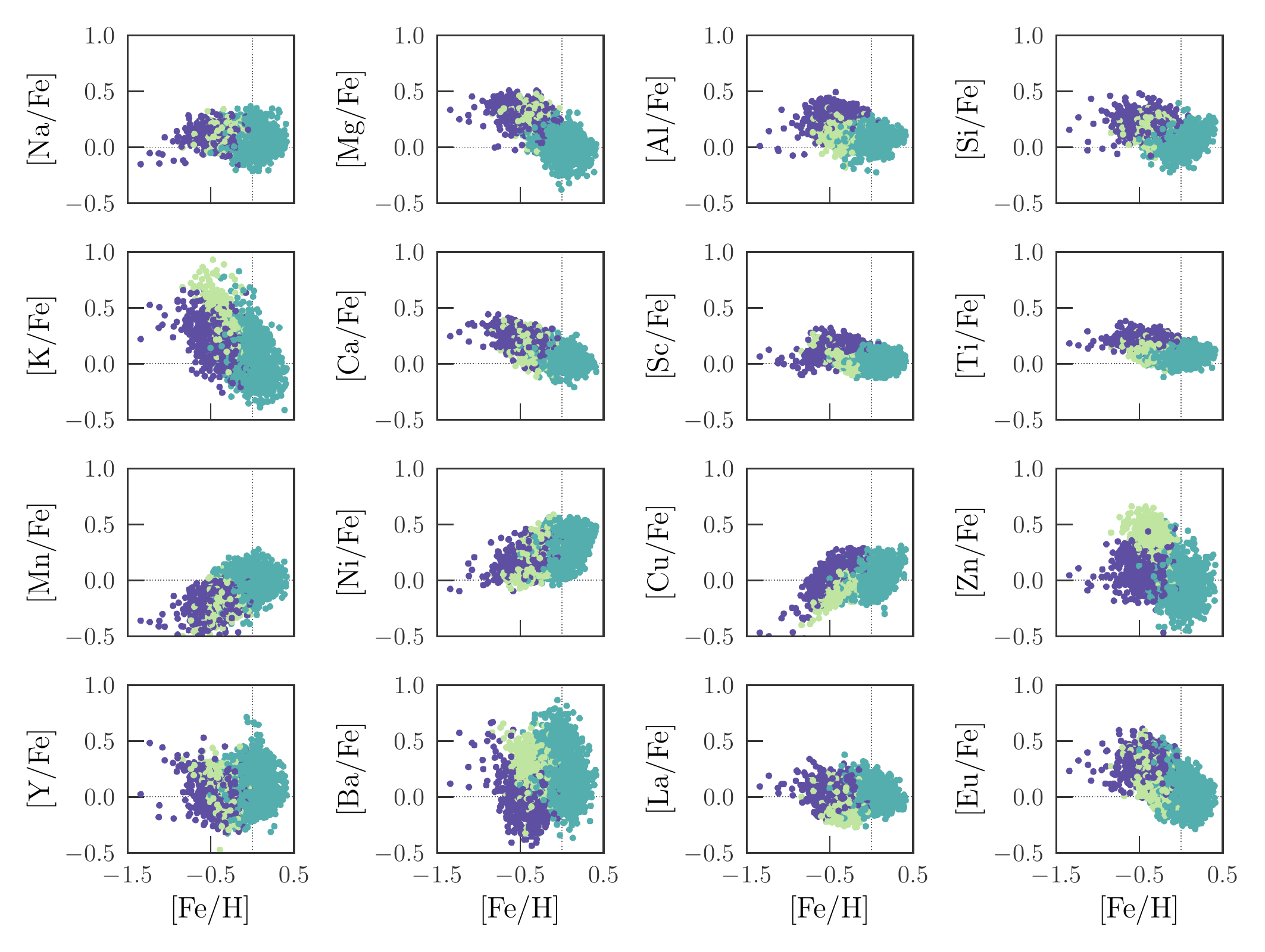}
	\caption{Detailed chemical abundances from the $N =$ \replaced{1,072}{2,566} stars in \Galah\ \citep{Buder:2018} that have 17 chemical abundances (Section~\ref{sec:exp4}). Each star is coloured by its $K$th inferred component from the lower-dimensional latent space, with the same colouring as per Figure~\ref{fig:exp3-latent-space}.}
    \label{fig:exp3-data-space}
\end{figure*}

In Figure~\ref{fig:exp3-latent-space} we show the inferred clustering in latent space,
where the separation between components is arguably best seen in the splitting between $\factorscores_6$ with respect to $\factorscores_2$ or $\factorscores_3$. When projected to data space (Figure~\ref{fig:exp3-data-space}) the third component (light green) is seen to have relatively higher abundance ratios of [K,Ba,Zn/Fe] at a given [Fe/H]. This is consistent with the clustering in latent space.

\subsection{\Galah\ survey data with an increasing number of stars with missing data entries} \label{sec:exp10}
Here we extend our experiment in Section~\ref{sec:exp4} to progressively
include more stars, even though those stars have some abundance measurements
missing. 
Specifically we started with the same subset of \replaced{1,072}{2,566} stars in Section~\ref{sec:exp4}
and added a random set of stars that met our criteria of \texttt{flag\_cannon = 0} and
\texttt{snr\_c1 > 30}.

\added{We initially added 1,000 stars to give a sample of $N =$ 3,566, then repeated
the grid search for the number of latent factors and components, and recorded
the model with the lowest message length. We then repeated this
procedure using 10,000 stars ($N =$ 12,566), and finally using all 
157,242 stars that met the criteria of \texttt{flag\_cannon = 0} and
\texttt{snr\_c1 > 30} to give a total sample size of $N =$ 159,808 stars.}

\begin{figure*}[t!]
	\centering
	\includegraphics[width=0.90\textwidth]{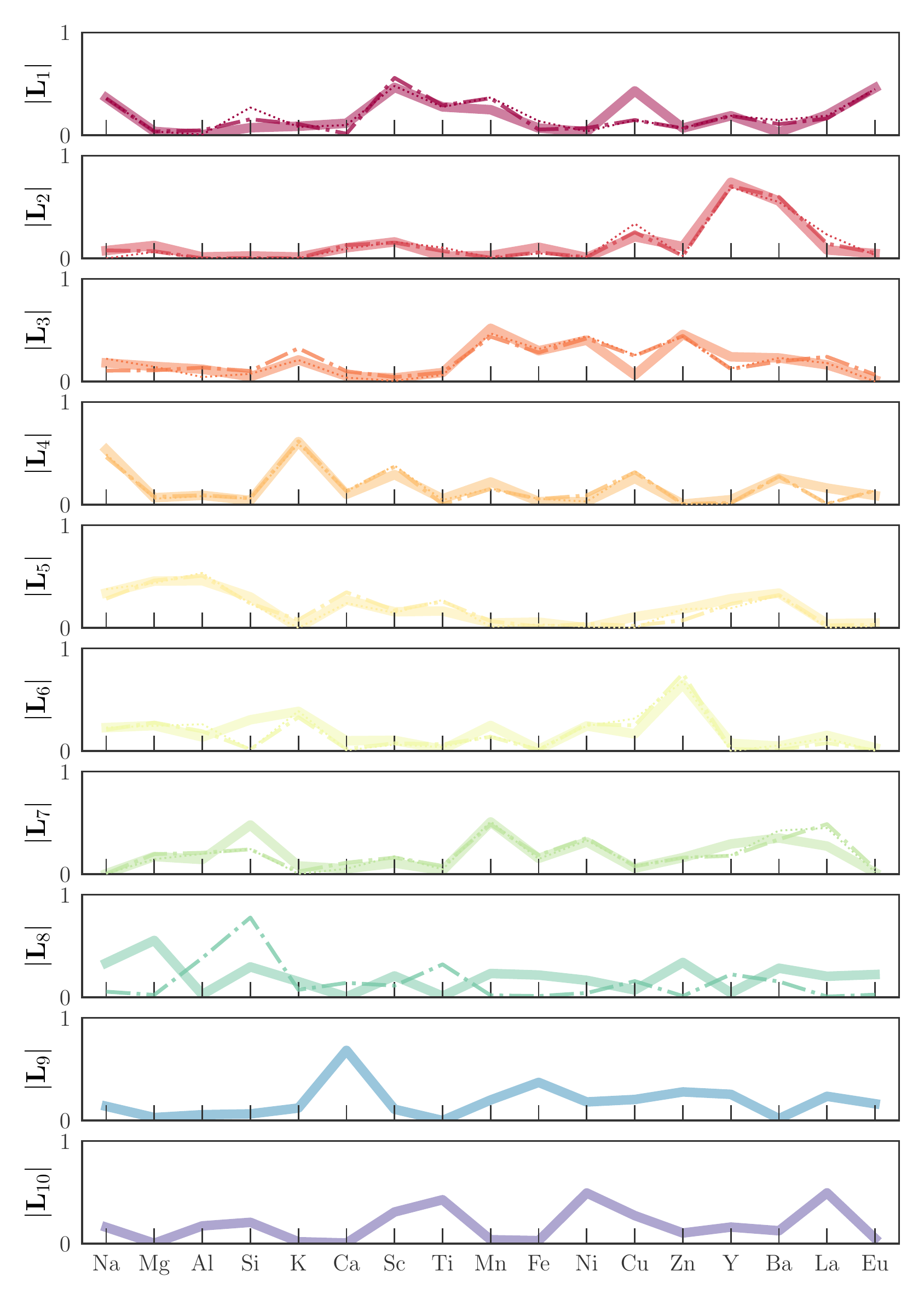}
	\caption{Latent factors found in Section~\ref{sec:exp10} using different subsets
	         of \Galah\ data \citep{Buder:2018}. The thin dotted line shows the result from
	         Section~\ref{sec:exp4} with $N=2,566$ stars with 17 abundances and no
	         missing data. The thicker dot-dashed line indicates $N = 12,566$ stars, of which $10,000$ have missing data entries. The thickest line has $N \sim 159,808$ stars, where ten latent factors are preferred.}
    \label{fig:exp10-comparison}
\end{figure*}

\added{For sample sizes up to $N \sim $3,566 we found that six latent factors were
preferred, and these factors shared common features (Figure~\ref{fig:exp10-comparison}).
This illustrates that the
first\footnote{`First' has no concept here in terms of factor load ordering, but
for the purposes of comparing inferred loads from different data sets we have ordered
the loads to be as close to those inferred in Section~\ref{sec:exp4}.} set of inferred factor
loads inferred from a smaller, complete data set, remain largely unchanged despite the increasing sample size and the increasing number of missing data entries.
When the sample size reaches $N = $12,566 we find another latent factor was 
required to best explain the data. When $N \sim$ 159,808, the preferred number of latent factors rises to ten ($J = 10$).}

\section{Discussion} \label{sec:discussion}

We have introduced a model to simultaneously account for the lower
effective dimensionality of chemical abundance space, and perform clustering
in that lower dimensional space. This provides a data-driven model of
nucleosynthesis yields and chemical tagging that allows us to simultaneously
estimate the latent factors that contribute to all stars, and cluster those 
stars by their relative contributions from each factor. The results are
encouraging in that we find latent factors that are representative of the
expected yields from dominant nucleosynthetic channels. However, the model that
we describe is very likely \emph{not} the correct model to use to represent 
chemical abundances of stars. Here we discuss the limitations of our model 
in detail.

We require latent factors to be mutually orthogonal in order to resolve
an indeterminacy. This suggests an astrophysical context where 
the mean nucleosynthetic yields (integrated over all stellar masses and star
formation histories) of various nucleosynthetic processes (e.g., $r$-process, 
$s$-process) are mutually orthogonal to each other. Clearly this assumption 
is likely to be incorrect: the nuclear physics
of one environment where elements are produced will be very different from
others, and there is no astrophysical constraint that those
yields (or latent factors) should be mutually orthogonal.
In principle one could represent the latent factors using a hierarchical data-driven model where the yields contribute as a function of stellar mass, 
metallicity, and other factors, but in principle to resolve the indeterminacy
\emph{in this model} would still require mutual orthogonality on the mean yields. Introducing a constraint on the factor scores that resolves this indeterminacy and allows for more flexible latent factors would be a worthy extension to this work.

The constraint of mutual orthogonality limits the inferences we want to make
about stellar nucleosynthetic yields. For example, after accounting for all
known sources of potassium production in the Milky Way, galactic chemical evolution
models under-predict the level of K in the Milky Way by more than an
order of magnitude \citep{Kobayashi:2006}. From our inferences 
using \Galah\ data, we find that $\factorloads_2$ -- the factor we identify as the s-process --  is the dominant contributor to
potassium. This latent factor persists even in the presence of missing data, and
a sample size two orders of magnitude larger.
Does this
suggest production of K is linked to the production of much heavier nuclei?
If our model could confidently and
reliably associate the production of K with other elements or sites then it could help explain the peculiar abundances of stars enhanced in
K and depleted in Mg \citep{Mucciarelli:2012,Cohen:2012} -- a chemical
abundance pattern that currently lacks explanation \citep{Iliadis:2016,Kemp:2018}.
In the \citet{Cohen:2012} sample their high [K/Fe] stars also tend
to be high in heavier elements, but there
are also numerous abundance correlations present.
However, is the K contribution that we infer physically realistic,
or is it a consequence of requiring that the latent factors are mutually
orthogonal? Distinguishing these possibilities is non-trivial, which is in
part why caution is warranted when trying to interpret latent factor models.
In this situation it is worth commenting that K has the largest specific
scatter (Figure~\ref{fig:exp3-specific-scatter}), suggesting that the contributions
of K are perhaps not as well described by the latent factor model as other
elements. This could in part be due to the non-trivial and significant effects
that the assumption of local thermodynamic equilibrium (LTE) has on our inferred K
abundances. These non-LTE effects are of order $0.5$~dex and will be accounted for in the upcoming \Galah\ data release (S. Buder, private communication).

A similar argument could be made for Sc, where $\mathbf{L}_1$ -- a factor load we
identify as the r-process -- is the primary contributor. Sc is under-produced in 
galactic chemical evolution models relative to observations \citep{Kobayashi:2006,Casey:2015}.
Based on this work, is the production of Sc linked to the production of heavy nuclei? 
Unlike K, the specific scatter in Sc is remarkably low: just 0.03~dex, among the
best-described elements after Ti and Fe (0.01~dex). This would suggest that the
latent factor model is a very good description for the production of Sc, but it
does not prove that it is \emph{the} description for the production of Sc.

There are other issues in our model that relate to our assumption of mutual
orthogonality. Even if nucleosynthetic yields were truely mutually orthogonal,
then the latent factors we infer are only \emph{identifiable} up until an
orthogonal basis. As we have seen in our experiments, the ordering and sign 
of the latent factors is not described \emph{a priori}. This is both a feature
and a bug: unrestricted ordering and signs allow for the model parameters to be
estimated more efficiently because they can freely rotate as the model
parameters are updated, but it does mean that we
must `assign' the latent factors we infer as being described by an astrophysical
process (e.g., the first latent factor is r-process). A more general limitation
of this is that the latent factors can be multiplied by some arbitrary rotation
matrix, leading to latent factor loads that are very different from what was
estimated by the model, but still lead to the exact same data (or log likelihood,
or Kullback-Liebler divergence, etc). As a consequence, we can only `identify'
latent factors up until this rotation. We have sought to address this by constructing
rotation matrices where the entries for each latent factor correspond to our expectations
from astrophysical processes (whilst remaining orthogonal), but here we are limited
by what astrophysical processes we are \emph{expecting} to find within the constraint
of being mutually orthogonal.

This in part constrains our ability to identify new nucleosynthetic processes. For example,
let us consider a hypothetical situation where we would only expect there to be four 
nucleosynthetic processes that predominately contribute to the observed \Galah\ abundances,
but in practice we found that the data are best explained with
five latent factors. We construct a rotation matrix where the first four latent factors
describe the nucleosynthetic processes we expect to find. What of the fifth latent
factor? We can constrain the possible values of the fifth latent factor conditioned on
the requirement that all factors remain mutually orthogonal, but one can imagine that
some (or perhaps many) elements have entries where the fifth latent factor can have
near-zero or zero entries. Even if the mean nucleosynthetic yields are mutually
orthogonal, there are scenarios that one can imagine where there is a limited amount
we can say with confidence about that new nucleosynthetic process \citep[see also][]{Milosavljevic:2018}.

\added{There are similar limitations that arise due to our assumption about the clustering in latent space.
There is no justified reason why the factor scores should be well-described by multivariate normal
distributions. If the \emph{true} underlying scores were not distributed as multivariate normals
then one can imagine similar outcomes when directly fitting data with a mixture of gaussian distributions:
additional components would be required to describe complex (non-gaussian) shapes in data space. 
This situation of model mismatch is more extreme when fitting only data rather than the model
described here because some of the data complexity will be described by the orthogonal latent
factors. However, qualitatively the picture is the same: when the \emph{true} underlying distribution
in factor scores are not described by multivariate normals, additional components will likely be
introduced in order to describe non-gaussian features.}

Notwithstanding these issues, we have shown that a latent factor model which allows for clustering in latent space can adequately
describe chemical abundance data. 
We find \replaced{five}{six} latent factors from a small subset of \Galah\ data with complete abundances, and
those latent factors can qualitatively be described within the context of astrophysical yields. 
Those latent factors are recovered in larger samples where the data are incomplete.
That did not have to be the case: the mutually orthogonal latent factors could be entirely 
different from our expectations such that they did not have to match our expectations of
nucleosynthetic yields. Indeed, the inferred factors -- even after a
valid rotation -- could have made no astrophysical sense whatsoever. For this reason it
is encouraging that there is some interpretability in the latent factors. Indeed, in the elements where we find surprising associations (e.g., Sc and K), these are elements where galactic chemical evolution models are most discrepant from observations, even after accounting for systematic errors in abundance measurements (e.g., violations to the assumption of local thermodynamic equilibrium).

In the subset of \Galah\ data with complete abundances we find that three components are
preferred. These components can be described as those with (1) low- and (2) high-[$\alpha$/Fe] abundance ratios, and another (3) primarily differing in K, Ba, and Zn abundances at a given [Fe/H] and [$\alpha$/Fe] abundance ratio. When we include \replaced{$\sim$100,000}{$\sim$160,000} stars with up to 17 abundances, and assume the incomplete abundances are missing at random, we find that 16 components in latent space are preferred to explain the data. By construction these  components
are structured in their chemical abundances because of the projection from the latent
space, and by extension of each component having similar chemistry, each component
occupies realistic locations in a Hertzsprung-Russell diagram. When we
project these component associations to the data space we find that none of the inferred
components are structured or coherent in their positions or motions. However, in this sample of stars there are no gravitationally bound clusters where a reasonable (e.g. $\sim$30) number of stars have been observed. Clearly, more data would help to resolve a higher number of components.

Perhaps it is not so discouraging that none of the inferred components are structured in their positions or motions because there are no gravitationally bound clusters in the data. But there is clearly more that can be done in chemical tagging. Some components we infer have stars with positions and galactic orbits that would imply that they cannot have formed in the same star cluster. In these situations there is likely significant value in including joint probabilities on whether two stars could be associated to the same star formation site based on their dynamic properties. Similarly, although stellar ages are historically difficult to estimate precisely, can this imprecise information help inform weak priors or probabilities of two stars having the same association? There is an incredible amount of dynamical information available from \project{Gaia}, particularly for stars in the \Galah\ survey, and weakly informative priors might be sufficient to help improve the granularity of chemical tagging without being overly constraining on the dynamical and star formation history we seek to infer.

\section{Conclusions} \label{sec:conclusions}

We have introduced a data-driven model of nucleosynthesis by incorporating 
latent factors that are common to all stars, and allowing for clustering in the
lower-dimensional latent space. This approach simultaneously allows us to efficiently
tag stars based on their chemical abundances, and to infer the contributions that are
common to all stars (e.g., nucleosynthetic yields). Experiments with generated data
demonstrate that MML is a useful principle for selecting the appropriate number of
latent factors and components. Experiments with \Galah\ data reveal latent factors
that are qualitatively and quantitatively similar to expected nucleosynthetic yields (e.g., products
from the $s$-process, $r$-process, et cetera). Interestingly we find that deviations from
expected yields occur in elements where observations and galactic chemical evolution models
are most discrepant (e.g., K, Sc). While we advise caution in directly interpreting
those latent factors as being nucleosynthetic yields, our
model does provide the first data-driven approach to nucleosynthesis and chemical tagging. We advocate that more data, and the inclusion of weakly informative priors -- joint probabilities using astrometry and a simplified model of the Milky Way -- would help in realising the full potential of chemical tagging.

\acknowledgements
We thank the anonymous referee for a detailed review.
We acknowledge support from the Australian Research Council
through Discovery Project DP160100637.
The \Galah\ survey is based on observations made at the Australian Astronomical Observatory, under programmes A/2013B/13,
A/2014A/25, A/2015A/19, A/2017A/18. We acknowledge the traditional owners of the land on which the AAT stands, the Gamilaraay
people, and pay our respects to elders past and present.
This research has made use of NASA's Astrophysics Data System.

\software{
	\package{Astropy} \citep{astropy:v1,astropy:v2},\,\,
    \package{IPython} \citep{ipython},\,\,
    \package{Jupyter Notebooks} \citep{jupyter-notebooks},\,\,
    \package{matplotlib} \citep{mpl},\,\,
    \package{numpy} \citep{numpy},\,\,
    \package{scipy} \citep{scipy},\,\,
    \package{TOPCAT} \citep{Taylor:2005}
}

\bibliographystyle{aasjournal}
\bibliography{mcfa}

\appendix

Documentation for the software that accompanies this paper is available at \url{https://mcfa.rtfd.io}. 
Below we provide code that generates ficticious data from a toy model and fits it.\\

\begin{minted}[mathescape,linenos,numbersep=5pt]{python}
import numpy as np
from mcfa import (mcfa, grid_search, mpl_utils, utils)

np.random.seed(42)

# A boolean variable to indicate whether to perform a grid search
do_grid_search = (np.random.uniform() > 0.5)

# Generate data
X, true_theta = utils.generate_data(n_samples=1000,
                                    n_features=15,
                                    n_components=10,
                                    n_latent_factors=5)

# The data, X, has shape (1000, 15)
assert X.shape == (1000, 15)

if do_grid_search:
    # Perform a grid search for the number of components and latent factors.
    J_trial = np.arange(1, 11) # trial from 1 to 10 latent factors 
    K_trial = np.arange(1, 21) # trial from 1 to 20 components
    J_grid, K_grid, converged, metrics = grid_search.grid_search(J_trial, K_trial, X)

    # Return the model with the smallest message length
    model = metrics["best_models"]["mml"]

else:
    # Or just fit the data given a number of components and latent factors.
    model = mcfa.MCFA(n_components=10, n_latent_factors=5)
    model.fit(X)

# Fitting quantities have a _ suffix
tau = model.tau_ # responsibility matrix
theta = dict(zip(model.parameter_names, model.theta_)) # model parameters

# Plot the factors.
fig_factors = mpl_utils.plot_factor_loads_and_contributions(model, X)

# Plot latent space.
fig_latent = mpl_utils.plot_latent_space(model, X)

# Plot data space, coloured by most probable component.
fig_data = mpl_utils.plot_data_space(model, X)
\end{minted}

\end{document}